\documentclass[aps,prx,notitlepage,superscriptaddress,twocolumn]{revtex4-1}
\usepackage{amsmath}
\usepackage{amsfonts}
\usepackage{graphicx}
\usepackage{caption}
\usepackage{subcaption}
\usepackage{color}
\usepackage{hyperref}
\usepackage{comment}
\usepackage{braket}
\usepackage{dsfont}
\usepackage{array}
\graphicspath{./Figs}
\newcommand{\ztwo}{{\mathbb{Z}_2}}
\newcommand{\av}[1]{\left\langle #1 \right\rangle}	
\newcommand{\Hm}{\mathcal{H}}
\newcommand{\unitx}{\hat{e}_x}
\newcommand{\unity}{\hat{e}_y}
\newcommand{\unittau}{\hat{e}_\tau}
\newcommand{\sr}{{\bf r}}
\newcommand{\bq}{{\bf q}}
\begin{document}

\title{Charged fermions coupled to $\ztwo$ gauge fields: Superfluidity, confinement and emergent Dirac fermions.}

\author{Snir Gazit}
\affiliation{Department of Physics, University of California, Berkeley, CA 94720, USA}
\author{Mohit Randeria}
\affiliation{Department of Physics, The Ohio State University, Columbus, OH 43210}
\author{ Ashvin Vishwanath}
\affiliation{Department of Physics, University of California, Berkeley, CA 94720, USA}
\affiliation{Department of Physics, Harvard University, Cambridge MA 02138, USA }

\date{\today}

\begin{abstract}
We consider a 2+1 dimensional model of charged fermions coupled to a $\mathbb{Z}_2$ gauge field,  and study the confinement transition in this regime.  To elucidate the phase diagram of this model, we introduce a method to handle the Gauss law constraint within sign problem free determinantal quantum Monte Carlo, at any charge density. For generic charge densities, $\mathbb{Z}_2$ gauge fluctuations mediate pairing and the ground state is a gapped superfluid. Superfluidity also appears in the confined phase. This is reminiscent of the BCS-BEC crossover, except that a true zero temperature transition occurs here,  with the maximum $T_c$ achieved near the transition. At half-filling also one obtains a large Fermi surface which is gapped at zero temperature. However, on increasing fermion hopping a  $\pi$-flux phase is spontaneously generated, with emergent Dirac fermions that are stable against pairing. In contrast to a Fermi liquid of electrons, the change in Fermi surface volumes of the $\mathbb{Z}_2$ fermions occurs without the breaking of translation symmetry.
Unexpectedly, the numerics indicate a single continuous transition between the  deconfined Dirac phase and the confined superfluid, in contrast to the naive expectation of a split transition, where a gap to fermions precedes confinement. 

\end{abstract}


\maketitle

\section {Introduction}

In recent years, gauge theories have increasingly appeared in the description of condensed matter systems.   In contrast to high energy physics, where the lattice gauge theory is an approach to regularize a continuum quantum field theory and analyze confinement~\cite{KogutDuality}, in condensed matter systems gauge fields are emergent and lattice gauge theories are effective low-energy theories.  
Important examples include the dual vortex theory of lattice bosons in two dimensions\cite{PsekinDuality,DasguptaDuality,FisherDuality}, theories of quantum antiferromagnets \cite{AndersonRVB,ReadSachdevCPNm1,AffleckFlux,SenthilZ2},  
, quantum dimer models \cite{KivelsonRVB,MoessnerDimerModels} and frustrated magnets \cite{balents_spin_2010,SachdevSPN,JalabertfrustratedIsing}.

The simplest, and historically the first, example of a lattice gauge theory 
is the Ising lattice gauge theory (ILGT) with a discrete $\mathbb{Z}_2$ local symmetry.
It was introduced \cite{WegnerIsingGaugeTheory} as a statistical mechanics model
that exhibits a phase transition without any symmetry breaking. The ILGT undergoes a 
zero temperature confinement/deconfinement phase transition that is in the 3D classical Ising model universality class.
This can be seen easily by establishing a duality between the $(2+1)$D ILGT and the 
2D transverse field Ising model~\cite{KogutDuality}. The deconfined phase of the IGLT is of great interest 
in condensed matter physics, since it is one of the simplest examples of a state with topological order 
which exhibits long range entanglement despite exponentially decaying correlations, and 
ground state degeneracy on manifolds with nontrivial topology \cite{WenSpinliquids,ReadRVB,KitaevToricCode}.

Coupling to dynamical matter fields , bosonic or fermionic, can have a 
dramatic effect on the phase diagram of pure gauge theories and have
led to new insights. Some notable examples are  the smooth 
evolution between the confining phases of lattice gauge theories and the Higgs phase obtained by condensing bosonic
matter fields~\cite{FradkinShenkarHiggs}, an emergent deconfined phase in compact 
QED3~\cite{HermeleDeconfined,KleinertConfinement}, a theory 
that is known to be confining in the absence of matter fields and the loss of asymptotic freedom in 
3+1D QCD with a sufficiently large number of flavors of fermions.
	
In this paper we ask the fundamental question of elucidating the phases and phase 
transitions of dynamical fermions coupled to a $\mathbb{Z}_2$ lattice gauge theory in $(2+1)$ dimensions.
Analytical approaches for lattice gauge theories are in general useful only in the strong and weak coupling limits,
and Quantum Monte Carlo (QMC) simulations are the only known way to bridge the gap between these limits.
However, with a finite density of fermions, Quantum Monte Carlo (QMC) simulations are usually plagued by the
fermion sign problem. Integrating out the fermions in the imaginary time functional integral leads in general to an
effective action with a fluctuating sign, or even worse, one that is complex, and leads to uncontrolled statistical errors in the QMC.

For fermions coupled to IGLT, however, we show that there is no sign-problem and describe a QMC algorithm that works 
for arbitrary chemical potential, provided there are an even number of fermion flavors. We work with two spin flavors 
$\uparrow$ and $\downarrow$. The absence of a sign problem allows us to work on large lattices and low temperature and obtain
unbiased results that are numerically exact up to statistical errors.

Our main results are as follows. We find that the phase diagrams are qualitatively different for
the chemical potential  $\mu \neq 0$ versus $\mu = 0$, which enforces half-filling on the square lattice
with nearest neighbor hoping.

\begin{figure}[t!]
		\includegraphics[scale=0.4]{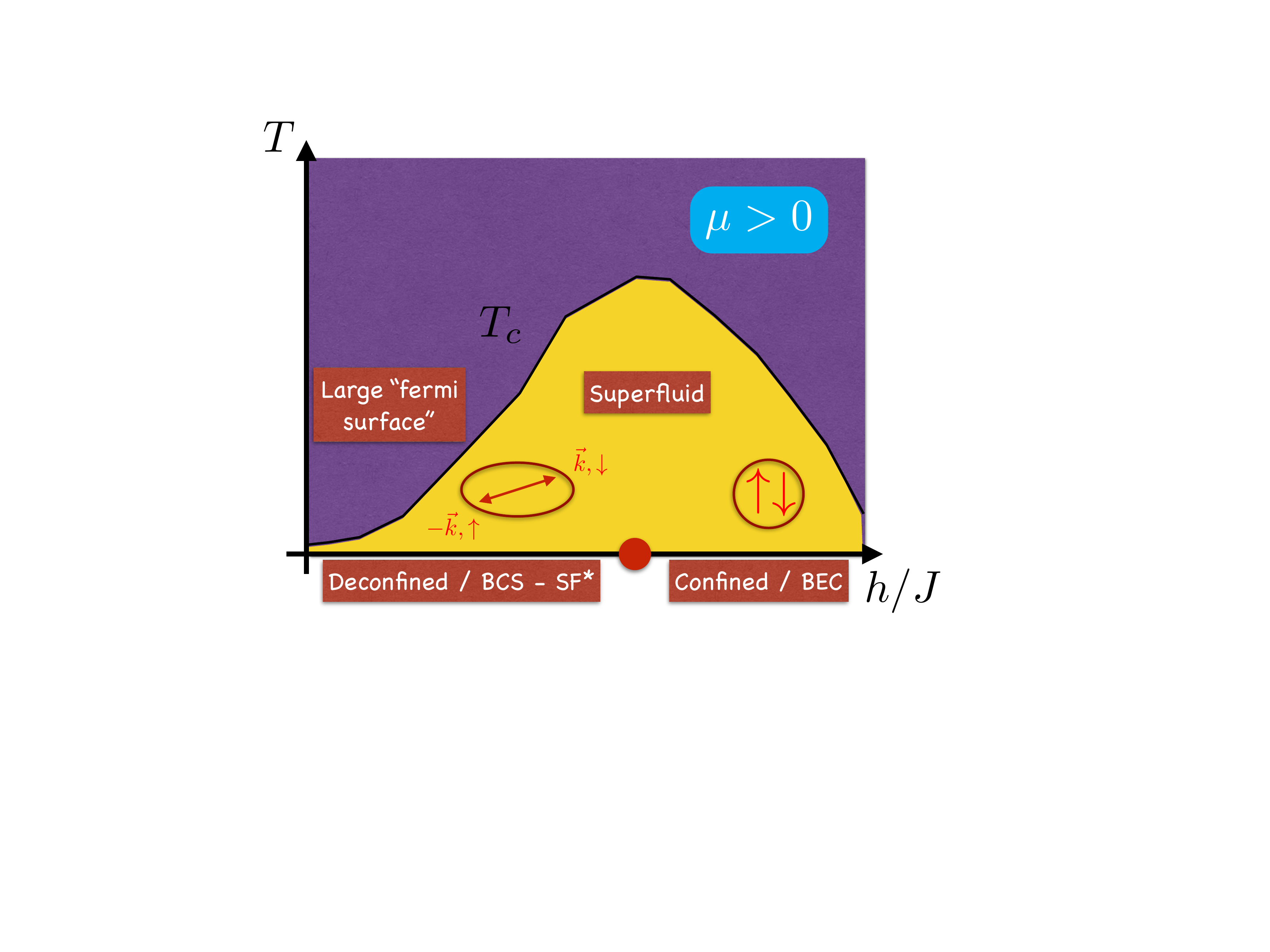}
		\caption{Schematic phase diagram at fixed $\mu>0$ as a function of $h/J$ and the temperature $T$.}
		\label{fig:phase_diag_mu_non_zero}
	\end{figure}

(1) For generic filling, we find that the $\mathbb{Z}_2$ gauge fields mediate an attractive interaction between the fermions, which are then
gapped due to pairing and form a s-wave superconducting ground state over the entire phase diagram. The superfluids in the opposite limits - deep in the deconfined phase and deep in the confined phase, are  reminiscent of the BCS and BEC scenarios respectively. However they differ at a fundamental topological level - in that the superfluid in the BCS side involves deconfined $\mathbb{Z}_2$ gauge fields and is an exotic superfluid (SF$^*$ in the notation of \cite{SenthilZ2} ). In contrast, a conventional superfluid is obtained on the BEC side, and hence the evolution from one to another cannot be via the usual 
 BCS to BEC crossover \cite{MohitBCSBEC} but must exhibit a zero temperature quantum phase transition at which the gauge fields undergo confinement (similar to the pure IGLT without fermions). This is shown in Figure \ref{fig:phase_diag_mu_non_zero} and discussed in Section \ref{sec:BCSBEC}.

\begin{figure}[t!]
		\includegraphics[scale=0.4]{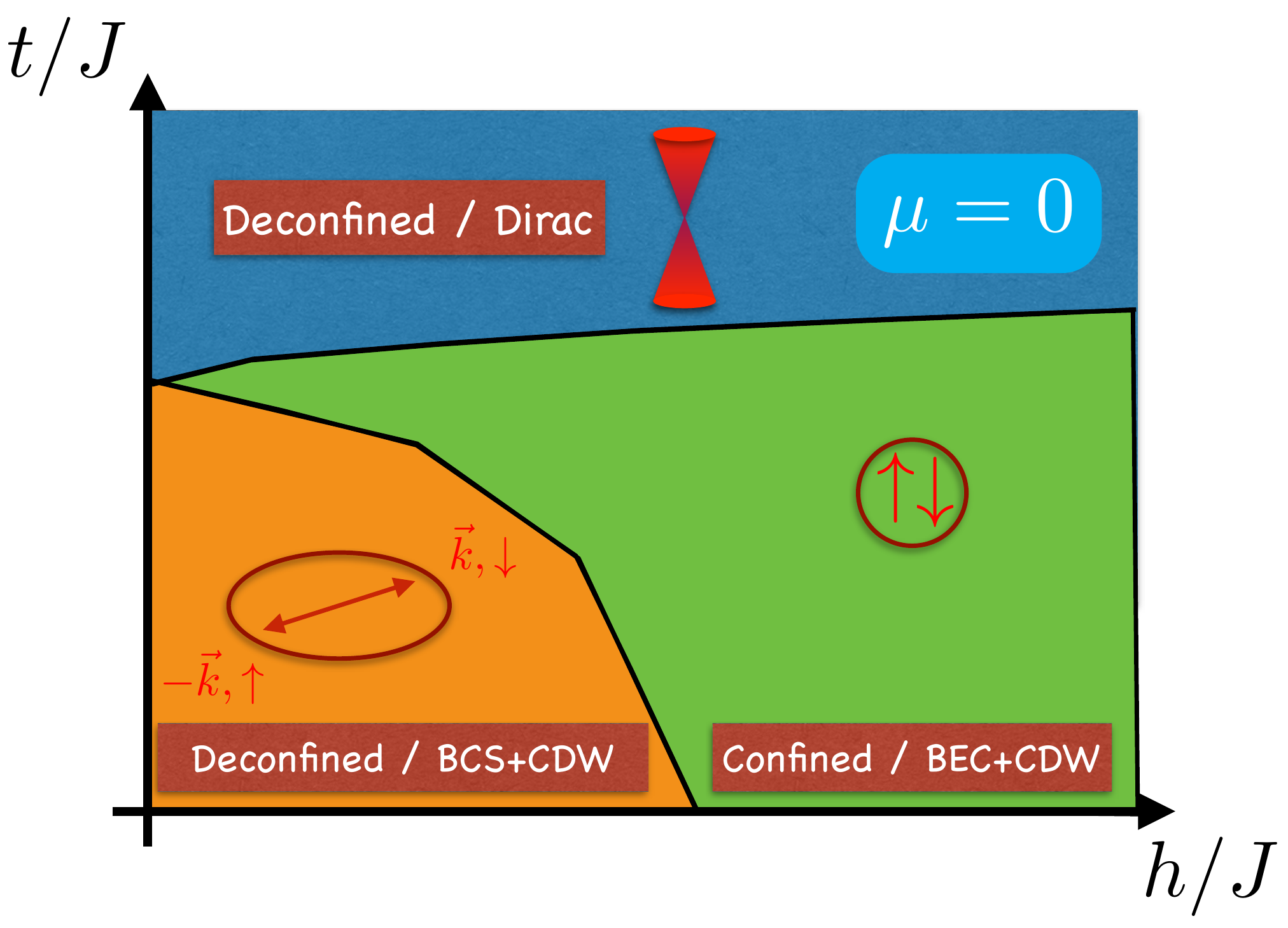}
		\caption{Schematic $T=0$ phase diagram at half filling with $\mu=0$.}
		\label{fig:phase_diag_mu_zero}
	\end{figure}

(2) Precisely at half filling, we find a new deconfined phase with emergent Dirac excitations, in addition to the 
deconfined BCS and confined BEC phases. We emphasize that we start with non-relativistic
fermions on a square lattice with near-neighbor hopping $t$. But, when $t$ is much larger than the ILGT coupling
constant, we find a spontaneously generated $\pi$-flux in every plaquette, which then leads to a Dirac excitation
spectrum. This is summarized in Figure  \ref{fig:phase_diag_mu_zero} and discussed in Section \ref{Sec:Dirac}.

	(3) We study the evolution of the deconfined Dirac excitations into the confined superfluid phase (BEC) (for example the vertical line in Fig. \ref{fig:phase_diag_mu_zero}) . Conventional wisdom holds that this would proceed through a split transition, wherein first the fermions would acquire a mass gap due to spontaneous breaking of symmetry (`chiral symmetry breaking' via a Gross-Neveu transition \cite{HerbutGross}), followed by a confinement transition in the usual Ising confinement universality class. Instead our numerics indicate a very surprising {\em single, continuous} transition wherein symmetry breaking and confinement occur simultaneously. The theoretical description of such a direct transition is an interesting open problem. Further numerical and theoretical study of this putative is left for the future but we summarize our current understanding in Section \ref{Sec:Dirac}. 

The $\mu=0$ results at weak coupling show a transition from a state with a 
large Fermi surface (the non-interacting limit of the deconfined BCS phase) to a deconfined Dirac phase
with point nodes. This is an amusing example of a change in Fermi surface area without a broken translational
symmetry that arises from the interactions of the fermions with gauge degrees of freedom.  
References \cite{SachecErezZ2,PunkDimer,NandkishoreMetltiskiSenthil} have suggested that related models and phenomenology may be  relevant to the study of strongly correlated electronic systems such as cuprates and heavy Fermion systems where Fermi volume changes appear to play an important role.

\section {Model and Methodology} 

\begin{figure}[t!]
  	\includegraphics[scale=0.5]{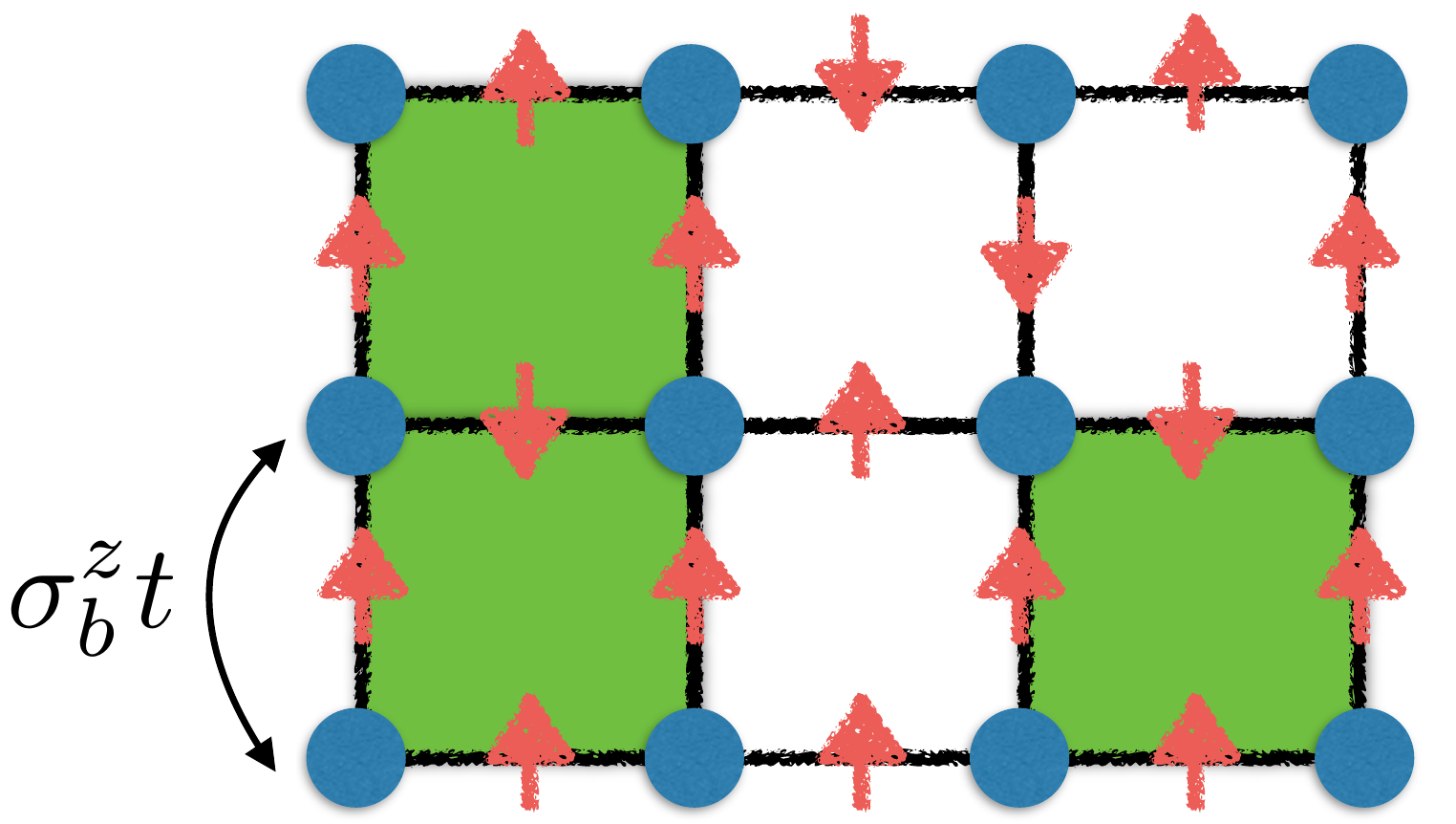}
  	\caption{Lattice model of Eq.~\eqref{eq:hamiltonian}: The Ising gauge fields $\sigma^z_b$ (red arrows) reside on the links of a square lattice. Frustrated plaquettes, with $\prod_{b\in \square_\sr} \sigma^z_{b}=-1$, are marked in green. The sign of the fermion hopping amplitude is determined by the Ising gauge field $\sigma^z_b$ along the bond $b$.    }
  	\label{fig:Z2LGT}
  \end{figure}

We consider the Hamiltonian \cite{SenthilZ2}
\begin{equation}
\Hm=\Hm_{\ztwo}+\Hm_f
\label{eq:hamiltonian}
\end{equation}
for the Ising lattice gauge theory (ILGT) coupled to fermions. The gauge degrees of freedom are  Pauli matrices
$\sigma^z_{r,\eta}$ and $\sigma^x_{r,\eta}$ residing on the bonds of a square lattice (see Fig.~\ref{fig:Z2LGT}) with site label
$\sr$ and $\eta=\unitx,\unity$. Their dynamics are governed by the Hamiltonian
\begin{equation}
\begin{aligned}
H_{\ztwo} = -J \sum_{\sr} \prod_{b\in \square_\sr} \sigma^z_b -h \sum_{\sr,\eta} \sigma^x_{\sr,\eta}
\end{aligned}
\end{equation} 
with a plaquette Ising magnetic flux term and bond electric field term.
The plaquette $\square_\sr$ is defined by the set of bonds  
$b\in\{(\sr,\unitx),(\sr,\unity),(\sr+\unitx,\unity),(\sr+\unity,\unitx)\}$.

\begin{figure}[t!]
 	\includegraphics[scale=0.35]{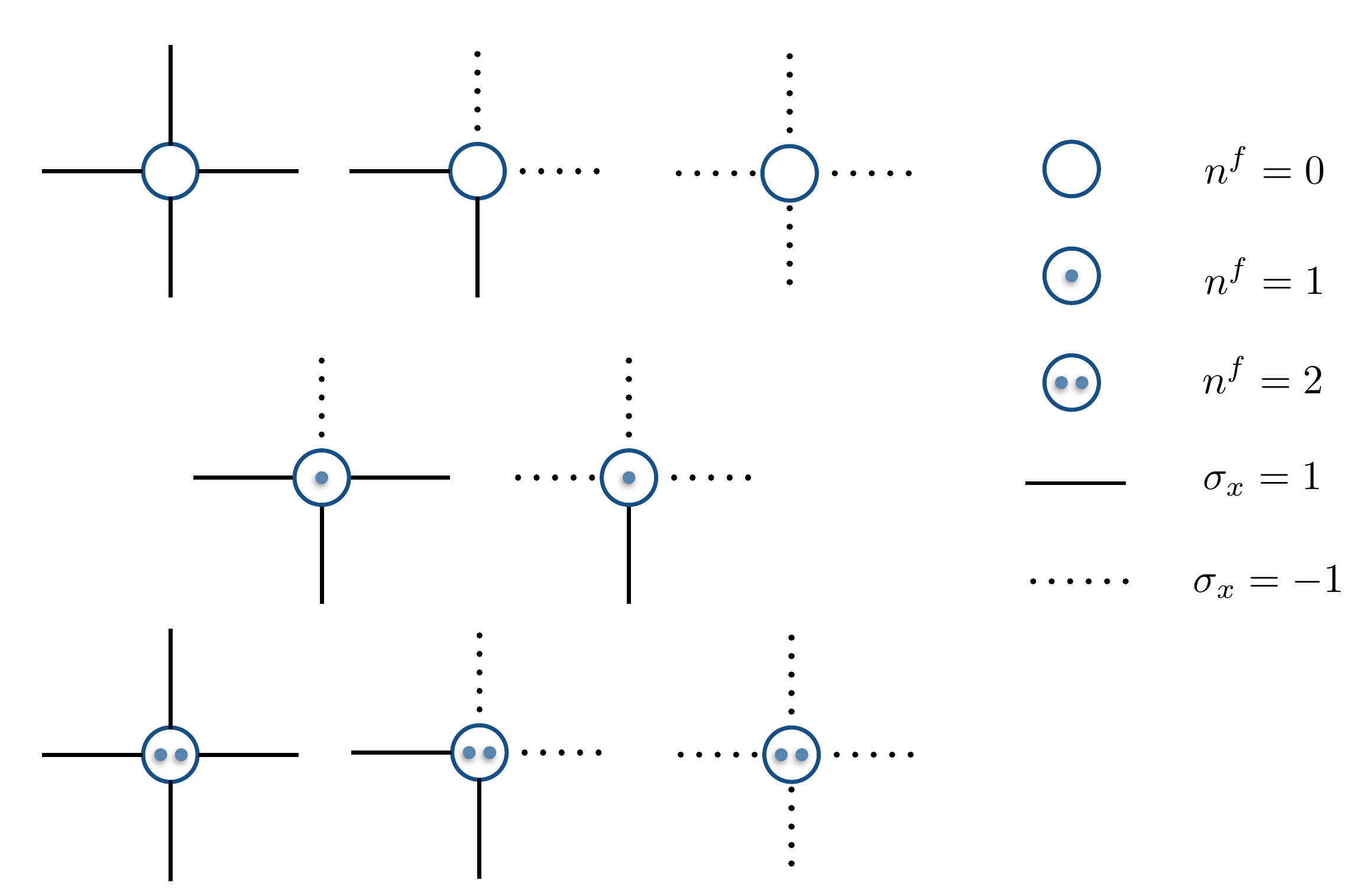}
 	\caption{Ising gauge field configurations in the $\sigma^x$ basis satisfying the constraint in Eq.~\eqref{eq:constraint} 
 		for different values of the fermion density $n^f$. Each fermion acts as a source of electric field.}
 	\label{fig:sigxconf}
 \end{figure}

The fermions hop between nearest neighbor sites and are minimally coupled to the gauge field though an Ising version of the Peierls substitution,
  \begin{equation}
  \Hm_{f} = -t\sum_{ \sr,\eta,\alpha} \sigma^z_{\sr,\eta}c^\dagger_{\sr,\alpha}c_{\sr+\eta,\alpha} + h.c. 
  -\mu\sum_{ \sr,\alpha}c^\dagger_{\sr,\alpha}c_{\sr,\alpha}.
  \end{equation} 
 Here $c^\dagger_{\sr,\alpha}$ is the fermion creation operator at site $\sr$ with spin $\alpha=\uparrow$ or $\downarrow$, 
 $t$ is the hopping amplitude and $\mu$ the chemical potential.
 We must restrict the Hilbert space to include only physical states that obey Gauss' law 
  \begin{equation}
\prod_{b\in +_\sr} \sigma^x_{b}(-1)^{n^f_\sr}=1,
 \label{eq:constraint}
  \end{equation} 
a local constraint at each site $\sr$.
  Here $n^f_\sr=\sum_{\alpha} c^\dagger_{\sr,\alpha}c_{\sr,\alpha}$ is the fermion number operator at site $\sr$ and $+_\sr$ is defined 
  by the set of bonds $\{b\}=\{(\sr,\pm\unitx),(\sr,\pm\unity)\}$, with $\sigma_{\sr,-\eta}=\sigma_{\sr-\eta,\eta}$. 
Each fermion acts a source for an electric field as shown in Fig.~\ref{fig:sigxconf}.
Eq.~\eqref{eq:constraint} restricts the Hilbert space to states without a background $\ztwo$ charge.

We impose the gauge-invariance constraint~\eqref{eq:constraint}
exactly via an Ising field that lives on temporal links, integrate out the fermions 
and show in Appendix \ref{app:DQMC} that there is no sign problem at any $\mu$.
We then sample the effective action for the gauge fields using standard determinantal QMC methods augmented by
global moves that are inspired by the worm algorithm; see Appendix \ref{app:worm} for details.

In addition to avoiding the sign-problem, we also need to introduce a technical innovation to 
circumvent the ``zero problem" that arises at the particle-hole (PH) symmetric point $\mu = 0$.
 We find that the vanishing probability for configurations odd under PH symmetry leads to a systematic bias 
in expectation values of observables that are not symmetric under PH transformation of a single spin flavor.
To address this problem, we introduce an extended configuration space that enables us to correctly sample these contributions
as shown in Appendix \ref{app:PH}.

Another challenge that we address below is the issue of characterizing various phases 
using only {\em gauge invariant} correlation functions. Commonly used
correlations like, e.g., the single particle Green's function are not gauge invariant and cannot be used
to characterize the excitation spectrum.

\section{Symmetries of the model}
Before discussing  the phase diagram of our model a few comments are in order. First, we note the crucial role played by global symmetries in defining our model (See Table.\ref{tab:table1}). We have implicitly assumed a global U(1) charge associated with the fermions, $c$, in addition to their $\mathbb{Z}_2$ gauge charge. For this reason we do not have explicit pairing terms in our Hamiltonian, and a Fermi surface is possible at least in the limit where gauge fluctuations are quenched. The charge U(1) symmetry is spontaneously broken in the  BEC/BCS superfluid phases.  

In the absence of a chemical potential $\mu=0$, the symmetry is enlarged and now includes a SU(2) pseudospin symmetry, just as in the Hubbard model\cite{JonesPseudoSpin}. The generators of this symmetry at site $\sr$ are: ${\bf P}_\sr ^{+} = (-1)^r c^\dagger_{\sr\uparrow}c^\dagger_{\sr\downarrow}$ and $P^z = (n_{\sr\uparrow} + n_{\sr\downarrow} -1)/2$, which commute with the $\mu=0$ Hamiltonian  (Eq
. \ref{eq:hamiltonian}).  Under this symmetry, superfluid order can be rotated into charge density wave order which must be degenerate at $\mu=0$. We always assume the SU(2) spin rotation symmetry is preserved. 

Another important symmetry is that of translations $T_x,\,T_y$, which allows us to define a unit cell and, in conjunction with the conserved U(1) charge, a filling. However, it is important to note that there is a degree of latitude involved in how the fermions $c$ transform, since they are not gauge invariant operators. That is , the symmetry action can perfectly well include a gauge transformation. 

The above can lead to  distinct symmetry  implementations, the projective symmetry groups\cite{Wenbook}, which respect the physical symmetry in all gauge invariant observables. In the context of translation symmetry, these are just the magnetic translation groups - which for the case of $\mathbb{Z}_2$ fluxes simply corresponds to the two translation generators of the square lattice, commuting or anticommuting i.e. $T_xT_y=\pm T_yT_x$ when acting on the fermions. 

These correspond to the zero and $\pi$ flux phases respectively, and at a fixed density of fermions would lead to a large or small Fermi surface whose volumes differ by a density corresponding to half filling. Nevertheless both these states are translationally symmetric, as can be seen by considering any observable (which is necessarily gauge invariant and hence blind to the flux). Note however, if these fermions were actually electrons, the phase with $\pi$ flux per unit cell would correspond to a doubling of the unit cell.   

Secondly, we would like to discuss the physical setting for the model described above and how it relates to more familiar condensed matter models. To do this, let us first determine the gauge invariant operators (See Table.\ref{tab:table2}), which are the physical degrees of freedom. 
This necessarily involves even powers of the fermion operator $c$, which can carry charge $2$  eg. $c^\dagger_\uparrow c^\dagger_\downarrow$ or charge $0$ eg. $c^\dagger c$ which in turn can either transform  as spin 1 or spin 0. 

The physical degrees of freedom in our model are readily identified deep in the confined phase when $h \rightarrow \infty$ in Eq. \ref{eq:hamiltonian}. There, we would like to set all the $\sigma^x=+1$. The constraint then implies that the fermion density on a  site $n^f_i =0 \,{\rm or}\, 2$, i.e. one has empty sites and sites occupied by a gauge neutral boson. In other words this is just the Hilbert space of a hardcore lattice boson model.  

On the other hand, if we had imposed the `odd' constraint - i.e. $\prod _{+_\sr}\sigma^x(-1)^{n^f_i} = -1$, then deep in the confined phase we would obtain $n^f_i=1$, and this could correspond to either a spin up or down fermion, which implies that we are dealing with a spin model. 

Our physical (gauge invariant) degrees of freedom then correspond to bosons with a global U(1) charge, and neutral spin excitations with integer spin. Therefore we are dealing with a boson only lattice model with spin and charge degrees of freedom.  Importantly there are no gauge invariant fermions - so this is not explicitly an electronic model. 

This is in contrast to the `orthogonal metal' to which our phase has many similarities (such as a Fermi surface of $\mathbb{Z}_2$ gauge charged fermions, carrying a global U(1) charge and spin 1/2 \cite{NandkishoreMetltiskiSenthil}) but differs in that the models of \cite{NandkishoreMetltiskiSenthil} explicitly involve electrons in the physical Hilbert space.  It would be an interesting exercise to reintroduce electrons as an additional degree of freedom to bring this closer to modeling correlated systems.

\begin{table*}[htp]
	\centering
	\caption{Symmetries of the gauge charged operators}
	\label{tab:table1}
	\begin{tabular}{p{5cm} p{4.5cm} p{3.5cm} p{4.5cm}}
		\toprule
 \bf Operator  \newline /Symmetry  & \bf Fermion: \newline$c_\alpha$    &\bf  Vison (`m' particle): \newline$v= \prod_{\rm string} \sigma^x$ &\bf  Boson (`e' particle): \newline$b_\alpha= c_\alpha v$\\
		\colrule
$U(1)$ charge                            & 1                                 & 0                              & 1                                      \\
		\colrule
		$SU(2)$ spin                            & 1/2                              & 0                             & 1/2                                    \\
		\colrule
	$T_xT_yT_x^{-1}T_y^{-1}$               & +1/-1, \newline(Large/Small Fermi Surface) & +1                             & +1/-1 ,\newline(Large/Small Fermi Surface)  \\
		\botrule
	\end{tabular}
\end{table*}
\begin{table*}[htp]
	\centering
	\caption{Symmetries of the gauge neutral (local) operators}
	\label{tab:table2}
	\begin{tabular}{p{5cm} p{4.5cm} p{3.5cm} p{4.5cm}}
		\toprule
		
		\bf Operator \newline / Symmetry & \bf Cooper pair: \newline$c_\uparrow c_\downarrow$    &\bf  Spin\newline {$c^\dagger_\alpha \sigma_{\alpha,\beta} c_\beta$}  &\bf  Energy Density: \newline$\prod_{\rm plaquette} \sigma^z$\\
		\colrule
		$U(1)$ charge                            & 2                                   & 0                              & 0                       \\
		\colrule
		$SU(2)$ spin                            & 0                               & 1                              & 0                                      \\
		\colrule
		$T_xT_yT_x^{-1}T_y^{-1}$               & +1 & +1                             & +1  \\
		\botrule
	\end{tabular}
\end{table*}

\section{Phase diagram}

\subsection{$\mu>0$: Confinement and BCS-BEC crossover}

\label{sec:BCSBEC}
Before describing the numerical results, we first establish the general structure of the phase diagram by considering several limiting cases.
This will also serve to summarize our main results. We note that the results are symmetric with respect to changing the sign of the hopping amplitude 
$t\to-t$, or the chemical potential $\mu\to-\mu$, or both. Without loss of generality, we consider 
$t>0$ and $\mu\ge0$. For simplicity, we focus on the case $J > 0$ and $h>0$. We choose the chemical potential to lie within 
the bandwidth $\mu \in [-8t,8t]$. In the following, we will distinguish between the cases $\mu>0$ and $\mu=0$ since, as we explain below, 
qualitatively different physics emerges at the half filling.

The phase diagram for $\mu>0$ is depicted schematically in Fig.~\ref{fig:phase_diag_mu_non_zero}.
We argue next that for a small, fixed value of $t/J$, one has a BCS-to-BEC crossover in the fermion sector 
with increasing $h/J$, together with the usual deconfined to confined phase transition
for the $\ztwo$ gauge fields. The QMC results are discussed in the following Section.

Deep in the confining phase,  $h\!\gg\!J$ and $h\!\gg\!t$, the ground state of the Ising gauge field sector 
is $\ket{{\rm GS}}_\text{{\rm confined}}=\prod_{\sr,\eta} \ket{\sigma^x_{\sr,\eta}=1}$. The fermions then form 
tightly bound bosonic molecules, or on-site $\ket{\uparrow\downarrow}$ pairs, in order to minimize the electric field cost 
necessary to satisfy the constraint~\eqref{eq:constraint}.
The molecule is described by the bosonic creation operator  $b^\dagger_i=c_{i,\downarrow}^\dagger c_{i,\uparrow}^\dagger$. 

Quantum corrections delocalize the bosons via a virtual process in which one of the constituent fermions hops to a neighboring site and the
other one follows. Gauss' law \eqref{eq:constraint} generates an electric field along the bond connecting the two sites, 
leading to an intermediate state with energy cost $h$ deep in the confined phase. The effective hopping amplitude for bosons
 is then $t_{\rm boson}\sim t^2/h$. In addition, there is an on-site hard-core repulsion between bosons arising from the 
 Pauli principle for the constituent fermions. The ground state in the confined limit is then a Bose--Einstein condensate (BEC)
 with  $\av{b_{{\bf q}=0}}\neq 0$.
  
Next we consider the limit $h \ll J$, deep in the deconfined phase. The (gauge invariant)
ground state of the Ising sector is a zero flux state, 
$\ket{{\rm GS}}_\text{{\rm deconfined}}=\prod_{\sr,\eta} \ket{B_{\square_\sr}=1}$, where
$B_{\square} = \sigma^z\sigma^z\sigma^z\sigma^z$ around the plaquette.
It is useful to work in the axial gauge with all $\sigma^z_{\sr,\unitx} = +1$ on an infinite lattice;
(note that this need not work on a cylinder or torus). The ground state then simplifies to
$\ket{{\rm GS}}_\text{{\rm deconfined}}=\prod_{\sr,\eta} \ket{\sigma^z_{\sr,\eta}=1}$.
In this gauge, the fermions decouple from the gauge fields in the $h=0$ limit, and their ground state
is the Fermi sea obtained by filling up the square lattice cosine band up to $\mu$.  

We expect~\cite{KogutDuality, FradkinBook} that for small $h/J$ 
the $\ztwo$ gauge fields meditate a weak, short-range attractive interaction between
the fermions. This would lead to Cooper instability of the  the Fermi surface
and a BCS-like ground state.  At $T=0$ this state is actually a fractionated superfluid, dubbed SF$^*$~\cite{SenthilZ2},
with vison excitations that lead to ground state degeneracy on a cylinder or torus.

In summary, the fermions are always gapped out by pairing and 
exhibit a BCS to BEC-like crossover as a function of increasing $h/J$, while the $\ztwo$ gauge fields exhibit
a deconfined to confined phase transition. Since the fermions are gapped,
the confinement transition should be in the same universality class as the pure $\ztwo$ gauge theory in $(2+1)$D,
which is dual to the transverse field Ising model.

There are several ways in which the BCS-BEC crossover in this model differs from previous studies of the crossover 
in lattice models like the attractive Hubbard model. Despite the fact that there is a smooth crossover at
finite temperatures, there is a $T=0$ phase transition 
between a fractionalized  BCS superfluid SF$^*$ and a BEC superfluid (SF). 
Second, the functional form of the effective interaction between fermions evolves with $h/J$
in a very interesting way. It is exponentially decaying in the deconfined phase, an attractive power-law at criticality~\cite{Peskin} and a 
linearly diverging potential in the confined phase. 
In the BEC regime, the bosons are not merely bound, but confined.
We will see some manifestations of this in the numerical results below.
Finally, in an attractive Hubbard model, say on a square lattice with near neighbor hoping, if one were to keep
the chemical potential $\mu$ fixed and keep increasing the attraction $|U|$ one will eventually go to an empty (or completely
filled) lattice depending on $\mu < 0$ (or $\mu > 0$). 

\subsection{$\mu=0$: Emergent Dirac phase} 
\label{Sec:Dirac}
 
The schematic phase diagram at half-filling ($\mu=0$) shown in Fig.~\ref{fig:phase_diag_mu_zero} is more interesting than
the case discussed above. In the weak hopping limit $t\ll J$, we do not expect a qualitative difference between $\mu=0$
and $\mu\neq 0$. One just obtains a BCS to BEC crossover in the fermion sector with
a  confinement transition in the gauge fields. 

On the other hand, the situation is qualitatively different for $t \gg J$.
The surprise here is the emergent Dirac phase for large hopping, that arises from the
spontaneous generation of a $\pi$-flux through each plaquette, as discussed in detail below.
An interesting consequence, in the $h = 0$ limit, is the evolution from a large Fermi surface at small $t/J$ to Dirac nodes at large $t/J$.
This is then an example of a change in the Fermi surface volume without any translational symmetry breaking. 
 
Let us first think about the large $t$ limit where the kinetic energy of fermions dominates all other terms in the Hamiltonian.
Determining the ground state then amounts to finding the Ising gauge field $\{\sigma^z_b\}$ configuration that minimizes 
the kinetic energy of the fermions. Exactly at half filling, the optimal gauge field configuration is a uniform $\pi$-flux phase, 
$\ket{\pi\text{-flux}}=\prod_{\sr,\eta} \ket{B_{\square_\sr}=-1}$  as was shown in \cite{FluxLieb,AffleckFlux,FluxArovas}. 
The dispersion relation of the $\pi$-flux lattice is a semi-metal \cite{AffleckFlux} with two distinct gapless Dirac nodes. 
The two Dirac nodes are not a consequence of spatial symmetry breaking and are, in fact, mandated by the 
Nielson-Ninomiya theorem.

The vanishing density of states of the Dirac nodes means that the pairing instability~\cite{BCSDiracKopnin,BCSDiracNandkishore} 
does not occur for arbitrarily small attraction (unlike a Fermi surface). This stabilizes the non-superconducting phase of 
deconfined Dirac fermions in the large hopping amplitude limit.

Consider the transition from the deconfined Dirac ($\pi$-flux) phase to the confined BEC 
at fixed, large $h$. This necessarily involves two distinct phase transitions:
a confinement transition of the ILGT and spontaneous $U(1)$ symmetry breaking associated superconductivity. 
In principle, these two transitions can occur separately leading to an intermediate
deconfined BCS phase. We will show below that we find no evidence of such an intermediate phase in our numerics,
which are consistent with a single transition between the deconfined Dirac and confined BEC phases.
In the scenario with a direct transition, the Dirac excitations are expected to play a role in the low energy physics 
and give rise to a new universality class of the confinement transition.

At small $h$, we find numerical evidence for two transitions as we go from 
a deconfined BCS state at small $t$ to a deconfined Dirac state at large $t$. We interpret the intervening phase a confined BEC state. To justify this, we note that since the $\pi$-flux lattice minimizes the kinetic energy of the fermions, increasing $t$ generates an effective magnetic plaquette term with a {\em negative } coupling $\tilde{J}<0$. As a result, the bare Ising magnetic flux coupling $J$ is renormalized to smaller values and for sufficiently reduced coupling a transition to a confined phase is expected.

The schematic $\mu=0$ phase diagram in Fig.~\ref{fig:phase_diag_mu_zero} reflects these expectations.

\section{Quantum Monte Carlo results}

The grand canonical partition function $\mathcal{Z}(\beta,\mu)$ at inverse temperature $\beta=1/T$ and chemical potential $\mu$ 
is defined as
\begin{equation}
\mathcal{Z}(\beta,\mu)=Tr\left[\hat{P} e^{-\beta (\mathcal{H}-\mu N)}\right].
\label{eq:partition_function}
\end{equation}
where the Trace is over both the $\mathbb{Z}_2$ gauge fields and the fermions.
$\hat{P}=\prod_\sr \hat{P}_\sr$ is a projection operator, with 
$\hat{P}_\sr$ enforcing the Gauss' law constraint~\eqref{eq:constraint} at each site $\sr$.
Note that each $\hat{P}_\sr$ commutes with $\mathcal{H}$ of eq.~\eqref{eq:hamiltonian}.
Expectation values are then defined as ,
\begin{equation}
\av{\mathcal{O}}=\frac{1}{\mathcal{Z}}Tr\left[\hat{P} e^{-\beta (\mathcal{H}-\mu N)}\mathcal{O}\right].
\label{eq:exp_value}
\end{equation}

We numerically sample the partition function~\eqref{eq:partition_function} and measure various observable 
using a determinantal QMC algorithm described in detail in Appendix~\ref{app:DQMC}. Importantly, we explain there how we 
impose the projection operators $\hat{P}_\sr$ and obtain a QMC algorithm free of the fermion sign problem
for arbitrary $\mu$. 

At the particle hole symmetric point, $\mu=0$, the QMC weight for a certain macroscopically large subclass of configurations vanishes. 
This in turn gives rise to a systematic bias of the Monte Carlo result. In Appendix \ref{app:PH} we provide a detailed description of this 
``zero problem" and  suggest a simple solution that we implement in our numerical calculation.

We discretize the imaginary time in steps of size $\Delta \tau$
satisfying $t \Delta \tau \le 0.125$, which we found to be sufficiently small to control the Trotter error.
Finally, we introduce a global updating scheme, inspired by the worm algorithm \cite{wormAlg}, which dramatically reduces 
the Monte Carlo correlation time, thus enabling us to simulate relatively large systems in close vicinity to the critical coupling. 
Further details can be found in Appendix  \ref{app:worm}  	

In the presence of a dynamical gauge fields, Elitzur's theorem implies that only gauge invariant 
observables have a non-vanishing expectation value. We now discuss the various observables that we 
have computed to characterize the various phases and phase transitions.

As a probe of the Ising gauge field sector we consider the average Ising magnetic flux energy $\left\langle B \right \rangle$

with
\begin{align}
B   = \sum_{\sr}\prod_{b\in \square_\sr} \sigma^z_b 
\end{align} 
Near the confinement transition the magnetic field energy is expected to develop a singularity that is captures by 
susceptibility,
\begin{equation}
\chi_B= \frac{1}{N}\int_0^\beta\!d\tau\left[ \av{B(\tau)B(0)} -\av{B(0)}^2 \right]
\label{eq:chi_B}
\end{equation}
where $N$ is the number of sites.

Superconducting order is probed by studying 
the s-wave pairing susceptibility, defined as,
\begin{equation}
P_{SC}\left(\sr-\sr',\tau\right)=\frac{1}{2} \av{b^\dagger_{\sr'}(\tau)b_{\sr}+h.c}
\end{equation} 
where $b^\dagger_\sr=c^\dagger_{\sr,\downarrow}c^\dagger_{\sr,\uparrow}$ is the pair creation operator.

The current response to an external probe $U(1)$ electromagnetic 
gauge field is given by~\cite{scalapinoInsMetal},
\begin{equation}
\Pi_{\mu,\nu}(\bq,i\omega_m)=\av{-K_{\mu}}\delta_{\mu,\nu}-\av{J_\mu\left(\bq,i\omega_m\right)J_\nu\left(-\bq,-i\omega_m\right)}
\end{equation}
where the Matsubara frequency $\omega_m = 2\pi m/\beta$ with $m \in \mathbb{Z}$.
The diamagnetic term is given by (minus) the fermion kinetic energy along the $\mu$ direction.
\begin{equation}
K_{\mu}=-t\sum_{ \sr,\alpha} \sigma^z_{\sr,\mu}c^\dagger_{\sr,\alpha}c_{\sr+\mu,\alpha} + h.c.
\end{equation}
The current operator $J_{\sr,\mu}$ is defined as
\begin{equation}
J_{\sr,\mu}=-it\sum_{ \sr,\alpha} \sigma^z_{\sr,\mu}c^\dagger_{\sr,\alpha}c_{\sr+\mu,\alpha} - h.c.
\end{equation} 
We are only interested in the static response here, which we decompose into its longitudinal (L) and transverse (T)
parts
\begin{equation}
\Pi_{\mu,\nu}(\bq,i\omega_m=0)= \Pi^L(\bq) {q_\mu q_\nu \over q^2} + \Pi^T(\bq) \left(\delta_{\mu,\nu} - {q_\mu q_\nu\over q^2}\right).
\end{equation} 

\begin{figure}	[t!]
	\begin{subfigure}[t]{0.23\textwidth}
		\includegraphics[scale=0.36]{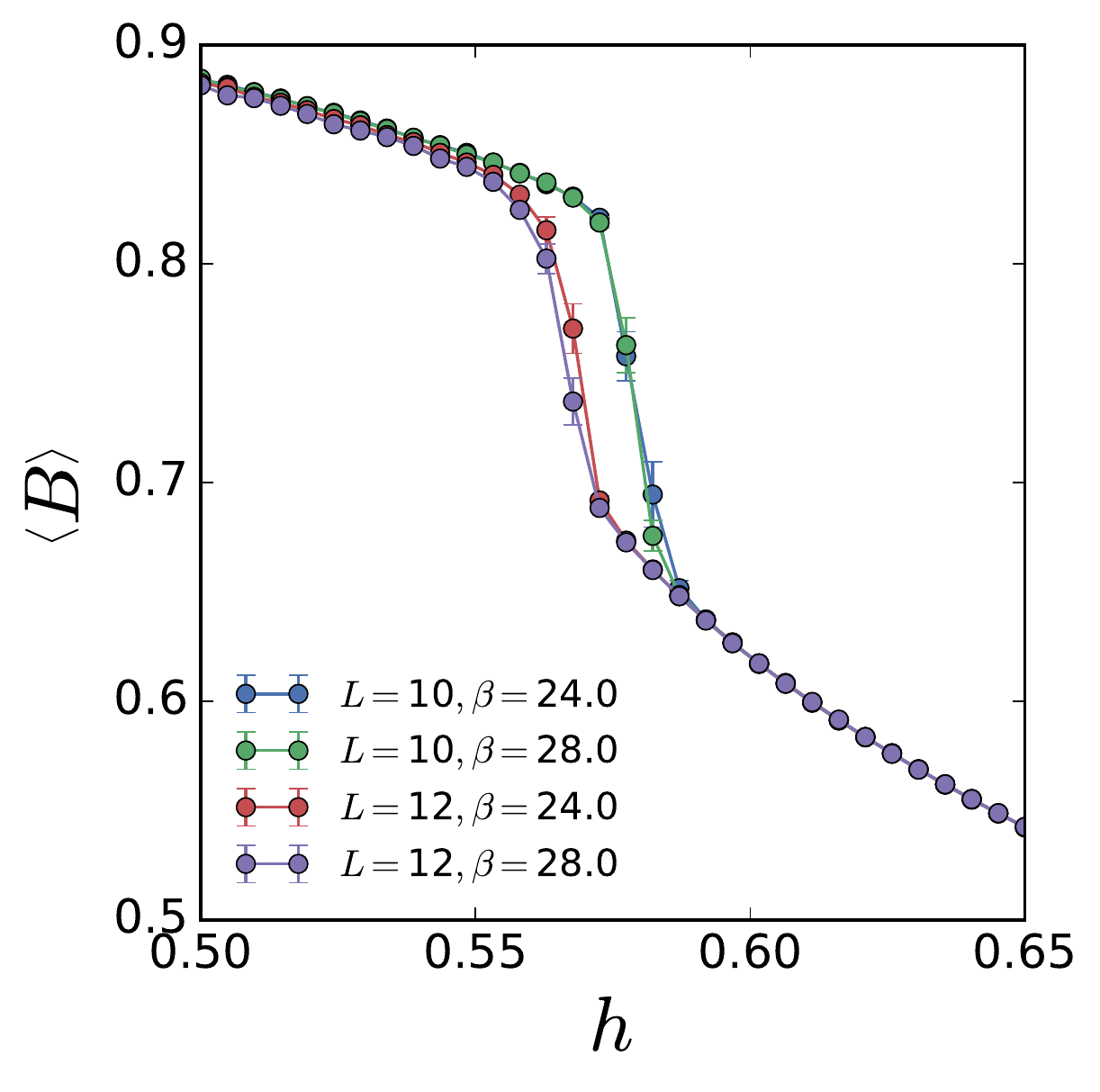}
		\caption{}\label{subfig:B_bcsbec}		
	\end{subfigure}
	\hfill
	\begin{subfigure}[t]{0.23\textwidth}
		\includegraphics[scale=0.36]{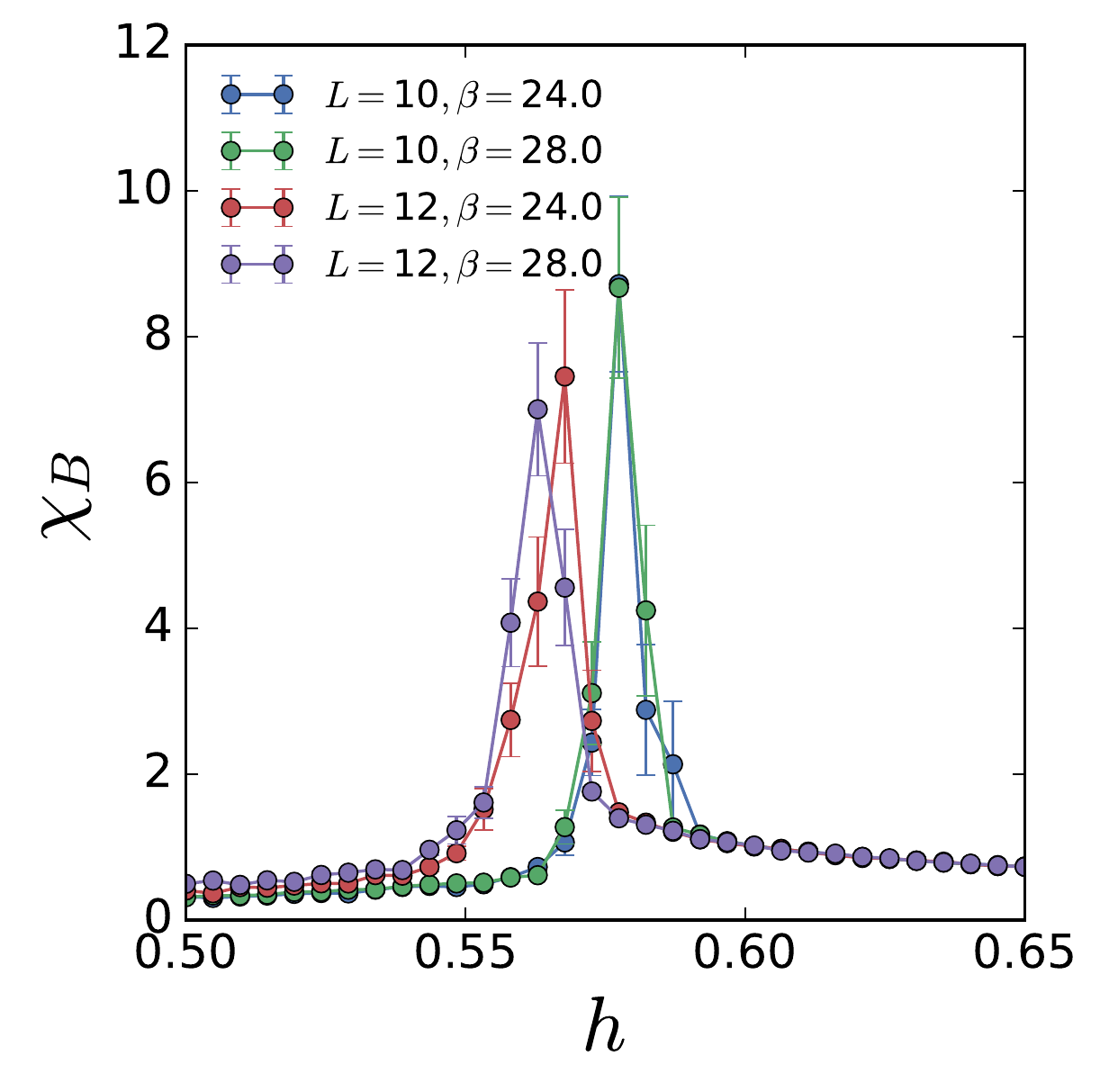}
		\caption{}\label{subfig:chiB_bcsbec}
	\end{subfigure}
	\caption{Confinement transition at $J=1, t=0.5, \mu=0.3$ as a function of  $h$. (a) The average Ising magnetic flux $\langle B \rangle$ drops from 
		unity deep in the deconfined phase at small $h$ to zero deep in the confined phase at large $h$. (b) Magnetic flux susceptibility $\chi_B$ 
		shows a peak at the confinement transition.}\label{fig:bcsbec}
\end{figure}
To characterize a superconducting state, we compute the superfluid stiffness 
$\rho_s=\Pi^T\left(\bq\to 0\right)$. In practice, on an $L \times L$ lattice, we compute~\cite{scalapinoInsMetal}
\begin{equation}
 \rho_s=\lim_{L\to\infty}\Pi_{xx}\left(q_x=0,q_y={2\pi}/{L};i\omega_m=0\right).
\end{equation}

Finally we need to identify observables to characterize the deconfined Dirac phase. 
The spectrum of fermionic excitations is not directly accessible to us since the 
the single particle Green's function is not a gauge invariant quantity.
We use the static (dia)magnetic susceptibility
\begin{equation}
\chi({\bf q}) = {\partial M({\bf q}) \over \partial B({\bf q})}  = -\ {1\over q^2}\ \Pi^T\left(\bq\right),
\end{equation}
which can be related to $\partial M/ \partial H$ using $B = H + 4\pi M$.
Note that this is the {\em only} place in the paper where we use $B$ to denote the external 
magnetic field, related to $U(1)$ electromagnetism, and {\em not} the $\mathbb{Z}_2$ magnetic field!

To identify the Dirac phase, we will exploit the 
characteristic $1/q$ divergence for the diamagnetic $\chi(q)$ arising from
point nodes. It was first first pointed out in the context of graphene~\cite{MagneticSuscpGrapheneAndo} that
\begin{equation}
\chi_{Dirac}({\bf q}) = -\ {g_s g_v v \over 16 q}
\label{eq:Dirac_chi}
\end{equation}
where $g_s$ ($g_v$) are the spin (valley) degeneracies, $v$ is the the Fermi velocity
and we use units with $e=c=\hbar=1$.
Results for various other observable such as the compressibility, spin susceptibility, 
s-wave pairing correlations, and Wilson loops will be presented in a later publication.

\subsection{Confinement and superconductivity for $\mu > 0$}

\begin{figure}	[b!]

	\begin{subfigure}[t]{0.23\textwidth}
		\includegraphics[scale=0.36]{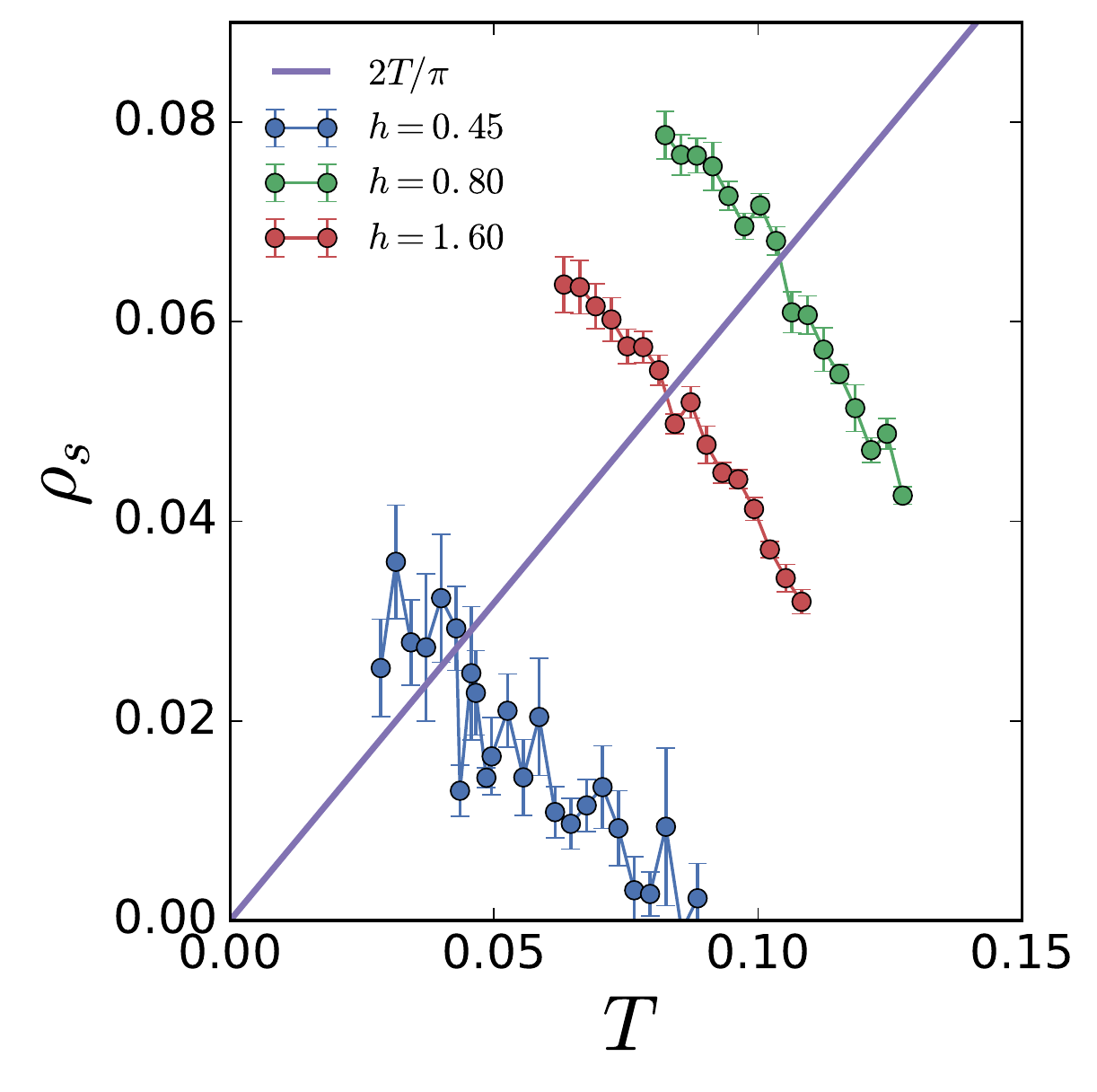}
		\caption{}\label{subfig:ktcross_bcsbec}			
	\end{subfigure}
	\begin{subfigure}[t]{0.23\textwidth}
		\includegraphics[scale=0.36]{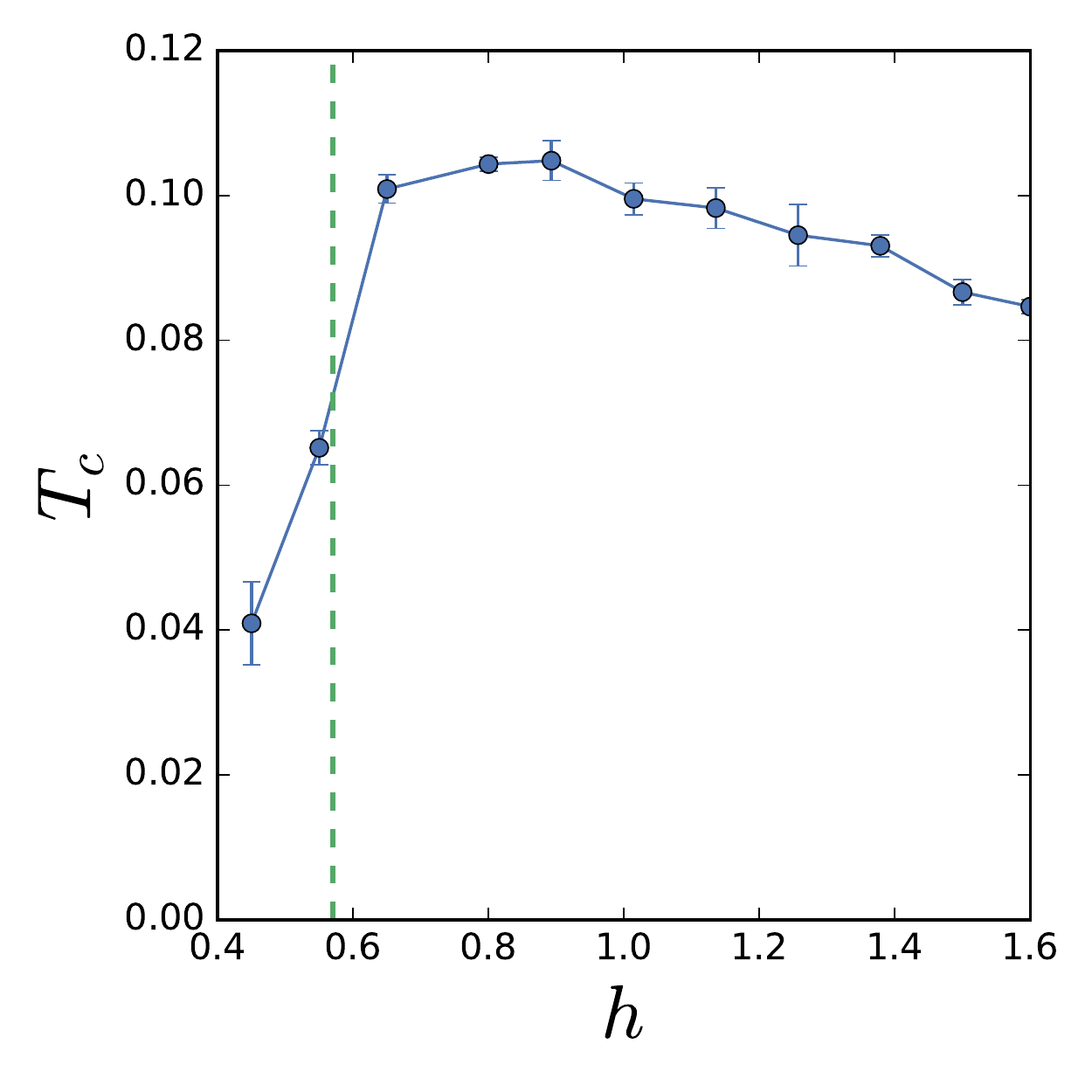}
		\caption{}\label{subfig:kt_bcsbec}	
	\end{subfigure}

	\caption{Superfluid stiffness $\rho_s$ and superconducting $T_c$ as functions of  $h$ for $J=1, t=0.5, \mu=0.3$ and $L=10$. 
		(a) Estimating the Berezinskii-Kosterlitz-Thouless $T_c$ from the universal jump in $\rho_s(T)$. 
		(b) Variation of $T_c$ across the confinement transition. The dashed line is the critical $h_c$ estimated from the peak position of $\chi_B$ in Fig.~\ref{subfig:chiB_bcsbec}, for $L=12$ and $\beta=28$.
		(c) Evolution of the low temperature $\rho_s$ across the confinement transition.}\label{fig:bcsbec}
\end{figure}

We now present QMC results for a $\mu > 0$ system which has a (large) Fermi surface in its noninteracting limit. Specifically,
we choose $J=1, t=1$ and $\mu=0.3$ and investigate the confinement transition and BCS-BEC crossover as a function of $h$.
In Fig.~\ref{subfig:B_bcsbec} we observe the following evolution of the average Ising magnetic flux $\av{B}$ with $h$. 
For small $h$, the plaquette term dominates in $\Hm$ and $\av{B}\to 1$. 
$\av{B}$ decreases with increasing $h$, since the electric field term generates quantum corrections 
in the form of $\pi$-flux vison excitations.  Deep in the confined phase, the electric field $E$ is frozen out and $B$
fluctuates wildly so that $\av{B}\to 0$ for large $h$.
We probe critical fluctuations near the confinement transition using
the susceptibility $\chi_B$ defined in eq.~\eqref{eq:chi_B}. We see in Fig.~\ref{subfig:chiB_bcsbec} 
that $\chi_B$ develops a peak that marks the confinement transition.
A quantitative analysis of the universal critical behavior requires a finite size scaling analysis, which will
be presented in a later publication.

We next show that the fermions are superconducting across the confinement transition. 
We see in Fig.~\ref{subfig:ktcross_bcsbec} that the system has a finite superfluid stiffness
$\rho_s(T)$ at low temperatures. We can estimate the transition temperature $T_c$ 
using the universal jump $\rho_s(T_c^{-}) =  2 T_c /\pi$, predicted by the 
Berezinskii, Kosterlitz Thouless (BKT) theory for 2D superconductors.
The finite size of our simulations, rounds off the jump discontinuity, but we
can nevertheless estimate $T_c$ from the intersection of the $\rho_s(T)$ curve with the
straight line $2/\pi T$ (see Fig.~\ref{subfig:ktcross_bcsbec}). 

The resulting $T_c$ estimates plotted in Fig.~\ref{subfig:kt_bcsbec} show that the 
transition temperature remains finite across the confinement transition at $h_c$ (marked with 
a dashed vertical line estimated from $\chi_B$). $T_c(h)$ has a non-monotonic variation 
with $h$ with a maximum that seems to be above the confinement $h_c$. Deep on the confined side,
we expect that $T_c$ (and $\rho_s(0)$) will both eventually vanish like $t^2/h$ in the BEC regime,
for reasons explained above. The very sudden drop in $T_c$ on the deconfined side
just below $h_c$, is related to the qualitative change in the attractive interaction 
between fermions (discussed in the previous section). This results in a small pairing gap and
thus a small $T_c$. Note that the energy gap cannot be estimated from the fermion Green's function, which is
gauge dependent. We are currently pursuing estimating the energy gap from the spin susceptibility ~\cite{RanderiaSpinGap,NandiniFL}.

\begin{figure}	[t!]
	\begin{subfigure}[t]{0.23\textwidth}
		\includegraphics[scale=0.35]{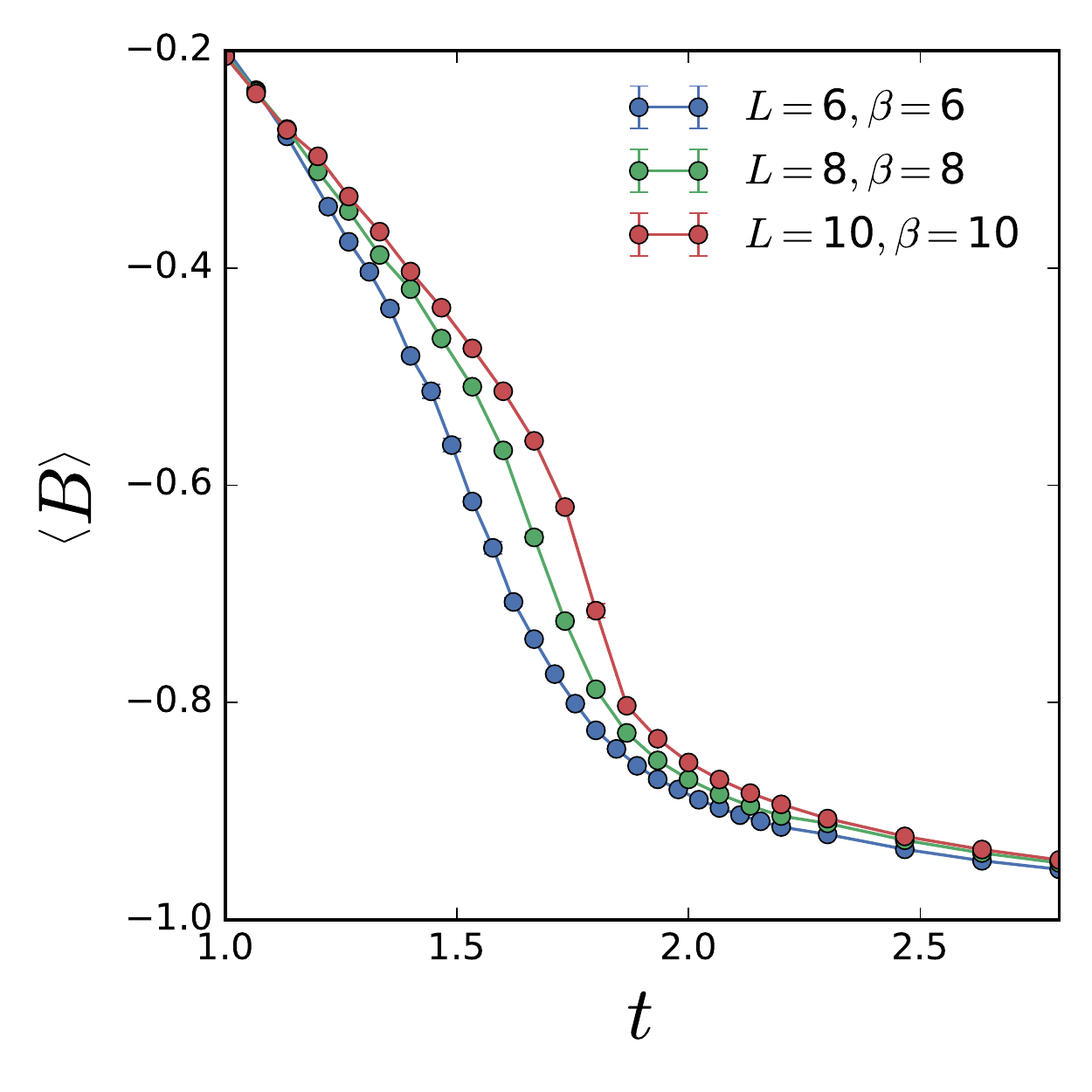}
		\caption{}\label{subfig:B_thop}		
	\end{subfigure}
	\hfill
	\begin{subfigure}[t]{0.23\textwidth}
		\includegraphics[scale=0.345]{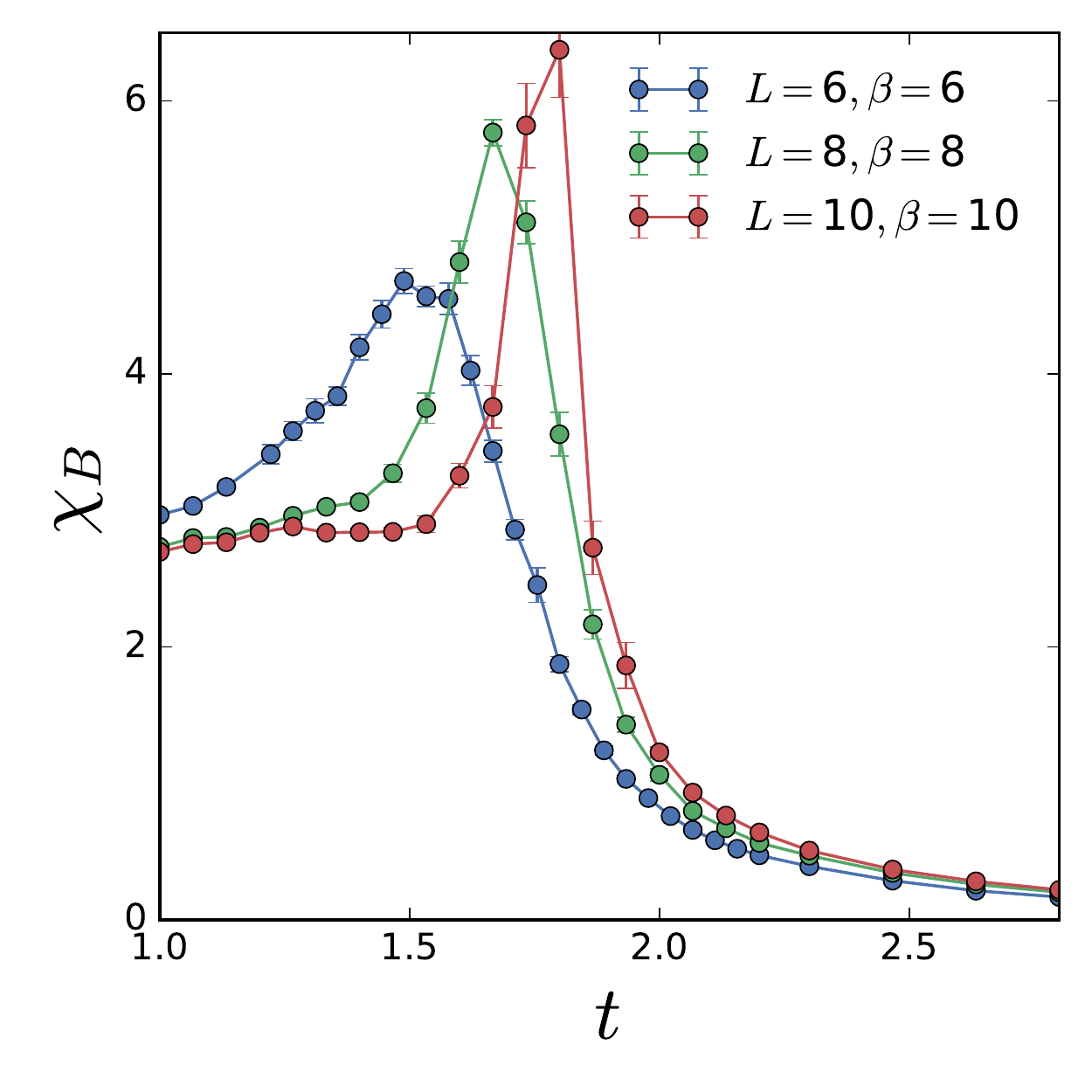}
		\caption{}\label{subfig:chiB_thop}
	\end{subfigure}
	
	\caption{Phase transition from a confined BEC to a deconfined Dirac phase driven by 
		by increasing $t$ at $\mu=0$ with $J=0.1,h=0.2$. (a) The average Ising magnetic flux $\langle B \rangle$ which goes from
		$0$ deep in the confined phase to $-1$ (``$\pi$-flux") deep in the deconfined limit. (b) The magnetic flux susceptibility $\chi_B$
		whose singularity indicates the confinement transition.}\label{fig:thop}	
\end{figure}

\subsection{Emergent Dirac excitations at $\mu = 0$}

\begin{figure}[b!]
	\begin{subfigure}[t]{0.23\textwidth}
		\includegraphics[scale=0.33]{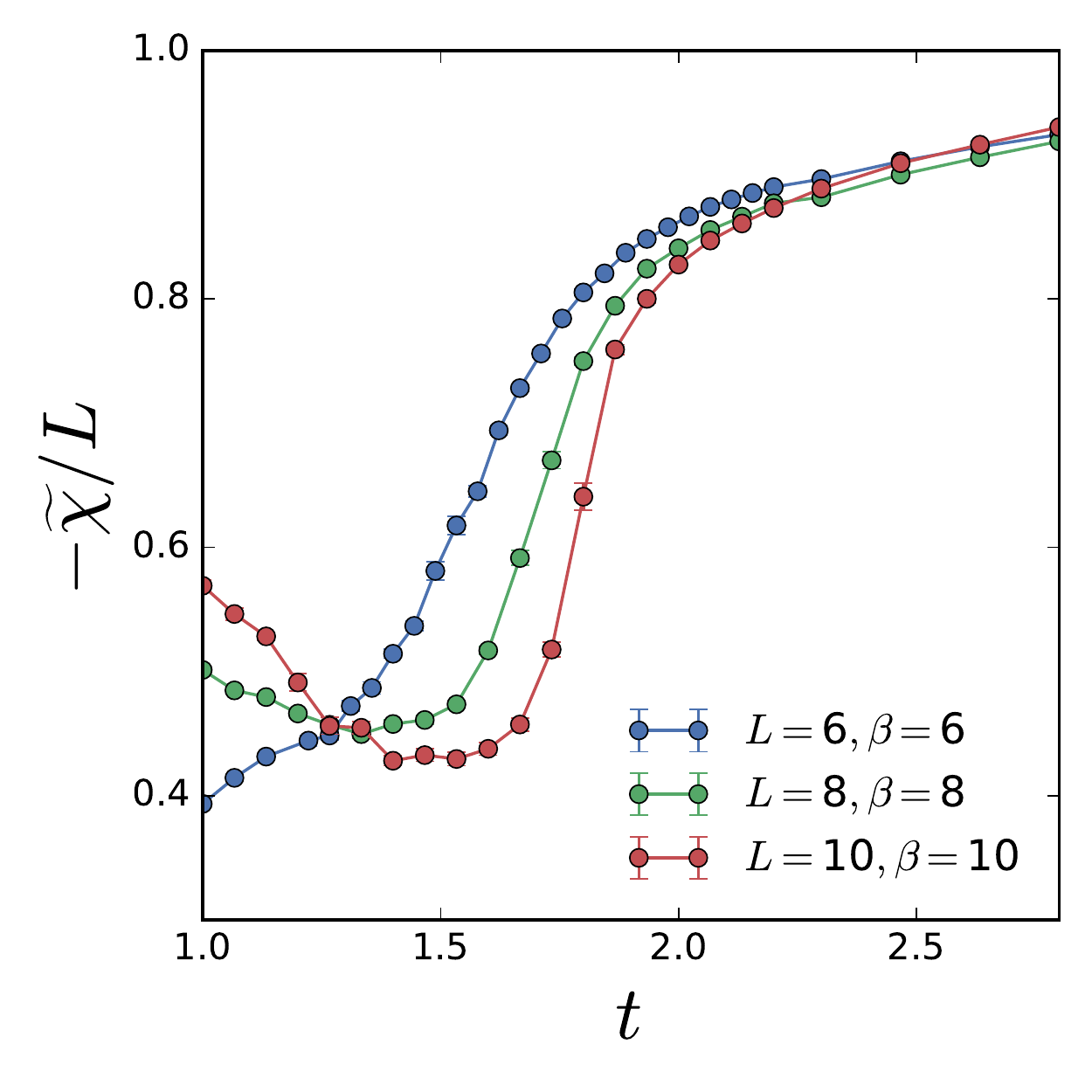}
		\caption{}\label{subfig:rhos_scling_thop}
		\hfill			
	\end{subfigure}		
	\begin{subfigure}[t]{0.23\textwidth}
		\includegraphics[scale=0.33]{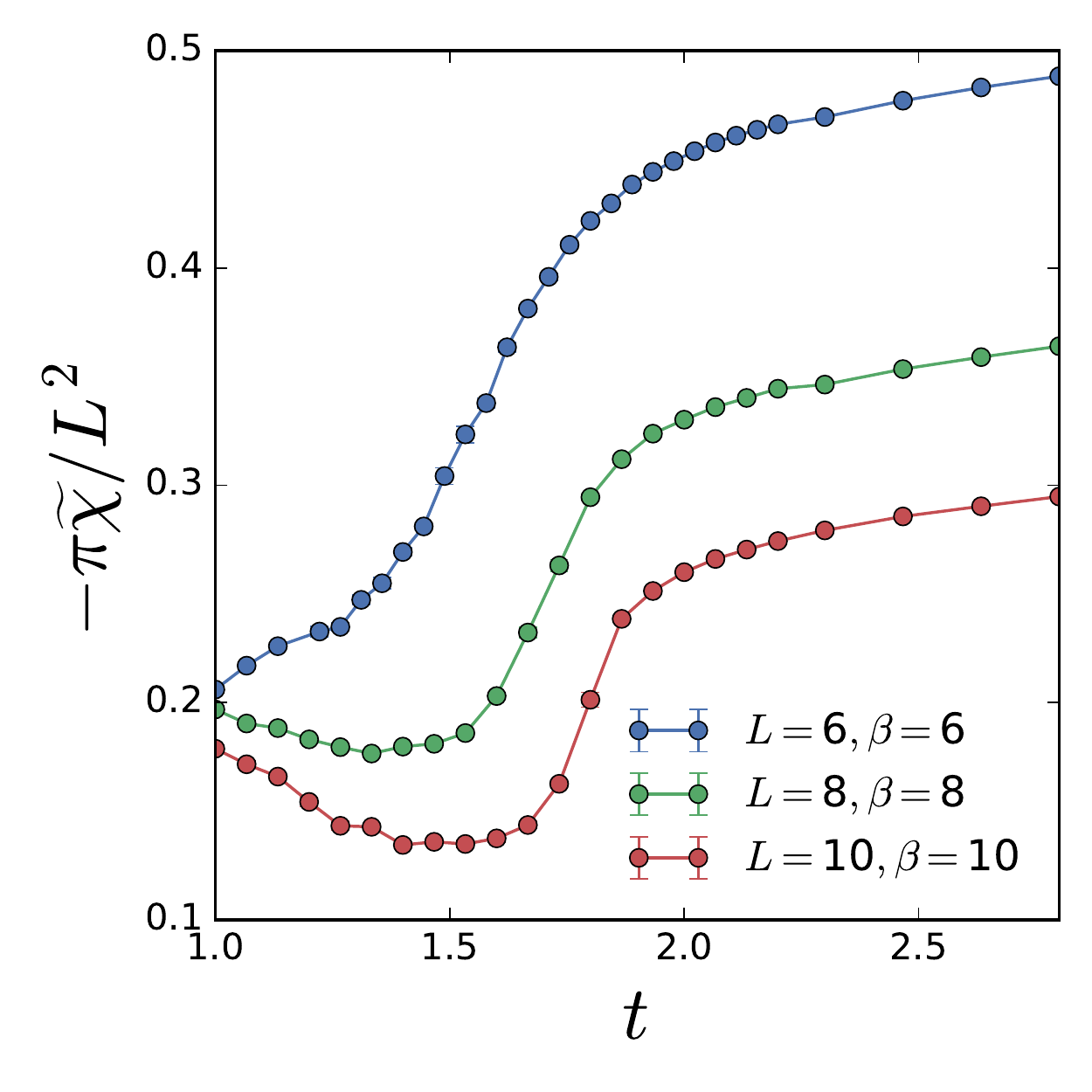}
		\caption{}\label{subfig:rhos_thop}	
	\end{subfigure}

\begin{subfigure}[t]{0.23\textwidth}
	\includegraphics[scale=0.33]{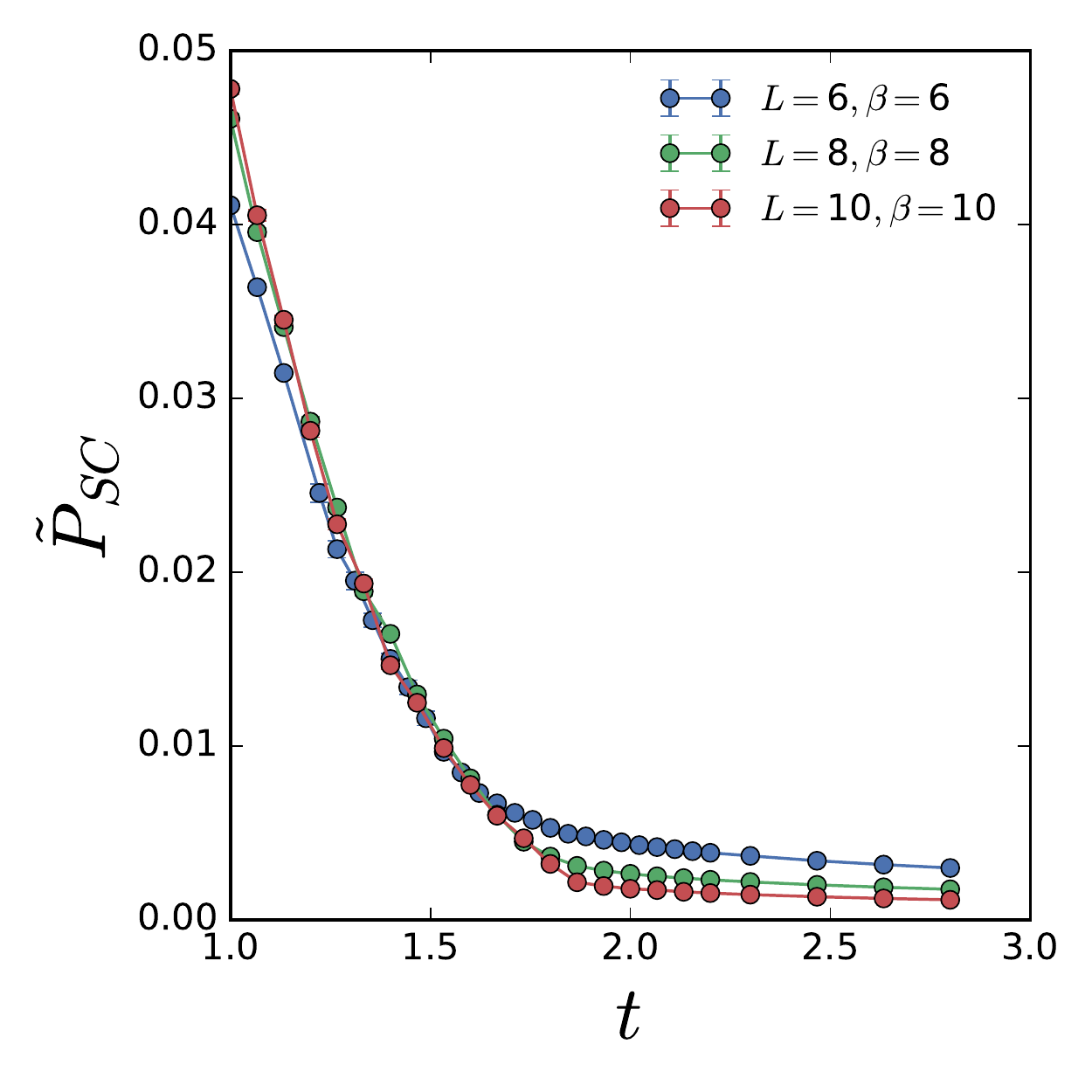}
	\caption{}\label{subfig:pairing_thop}	
\end{subfigure}

	\caption{	
		Confined BEC to deconfined Dirac phase transition at $\mu=0$ driven by varying $t$ for fixed $J=0.1, h=0.2$.
		The diamagnetic susceptibility $\widetilde{\chi} \equiv 4\pi\chi(q= 2\pi/L)/t$
		serves as a probe of both superconductivity and Dirac nodes.
		(a) $- \widetilde{\chi}/L$ as a function of $t$. This goes to unity independent of $L$ in the Dirac phase at large $t$; see text.
		(b) $- \pi\widetilde{\chi}/L^2$ as a function of $t$. Note that this goes to $\rho_s$, a constant independent of $L$ and $\beta$, 
		in the superconducting state at small $t$. 
		(c) We also study  superconductivity using the static zero momentum s-wave  paring susceptibility $\tilde{P}_{SC}$. In the confined phase the pairing susceptibility is finite indicating on superconducting order. With increase in $t$ the pairing susceptibility decreases and vanishes continuously at the critical coupling  $t_c=1.8(1)$. In the Dirac phase $\tilde{P}_{SC}$ remains zero.
}
\end{figure}

We finally turn to the particle-hole symmetric case $\mu=0$, where we expect an emergent Dirac phase based on the arguments presented in the previous section. We now present unequivocal numerical evidence for this interesting phase with excitations that become
gapless at nodal points in the Brillouin zone, even though we started with fermions with a simple non-relativistic dispersion.
For finite size and temperature scaling, we consider a sequence with increasing system size $L=6,8,10$ and 
set the inverse temperature to $\beta=L$. 

{\bf Strong Coupling:} 
First we investigate the transition at $\mu=0$ driven by varying $t$ at fixed $h/J > 1$, i.e., the strong coupling regime for the gauge theory.
We choose $J=0.1,h=0.2$ for the numerics.

The evolution of the average Ising magnetic flux $\av{B}$ with $t$ is plotted in Fig.\ref{subfig:B_thop}. 
In the small $t$ limit, deep in the confined phase, $\av{B} \to 0$. With increasing $t$ we see that $\av{B}$
decreases monotonically and asymptotically approaches $\av{B}\to-1$ in the large $t$ limit.
Thus a flux of $\pi$ per placate is spontaneously generated at large $t$.
We note that $J>0$ magnetic term in $\Hm$ favors zero flux (or $\av{B}= 1$) while
the $h > 0$ electric term in $\Hm$ randomizes the flux ($\av{B} = 0$). So the only way
to get a $\pi$-flux phase is if the fermion kinetic energy dominates both $J$ and $h$, 
together with $\mu$ giving rise to a commensuration effect.

To locate the the deconfinement transition we examine the susceptibility $\chi_B$. We see in Fig.~\ref{subfig:chiB_thop}. 
that $\chi_B$ displays a peak which increases with system size and marks the confinement (small $t$) to deconfinement 
(large $t$) phase transition at a critical value of $t_c\approx1.8(1)$. 
The universality class of this confinement transition is expected to quite different from that in the pure $\mathbb{Z}_2$ gauge theory, given that there are gapless Dirac excitations in the deconfined phase, as we show next.

We expect Dirac excitations in the $\pi$-flux phase, however, probing these is not so simple 
since quantities like the fermionic spectral function or density of states are not gauge invariant.
We therefore turn to the static diamagnetic susceptibility $\chi(q)$ in the long wavelength limit with
$q = 2\pi/L$.
In a Dirac phase with two spins ($g_s = 2$) and two nodes per Brillouin zone ($g_v = 2$),
eq.~\eqref{eq:Dirac_chi} implies $\chi_{Dirac}(q=2\pi/L) = -\ {v L/ 8 \pi}$, where $v$ is the Fermi velocity.
Thus $-4\pi\chi_{Dirac}(q=2\pi/L)/Lt = v/2t$, whose normalization has been chosen so that the answer is unity
for the non-interacting $\pi$-flux phase where the Fermi velocity at the Dirac nodes is  $v=2t$ as
sown in Appendix \ref{app:piflux}. 
\begin{figure}	[t]
	\begin{subfigure}[t]{0.23\textwidth}
		\includegraphics[scale=0.35]{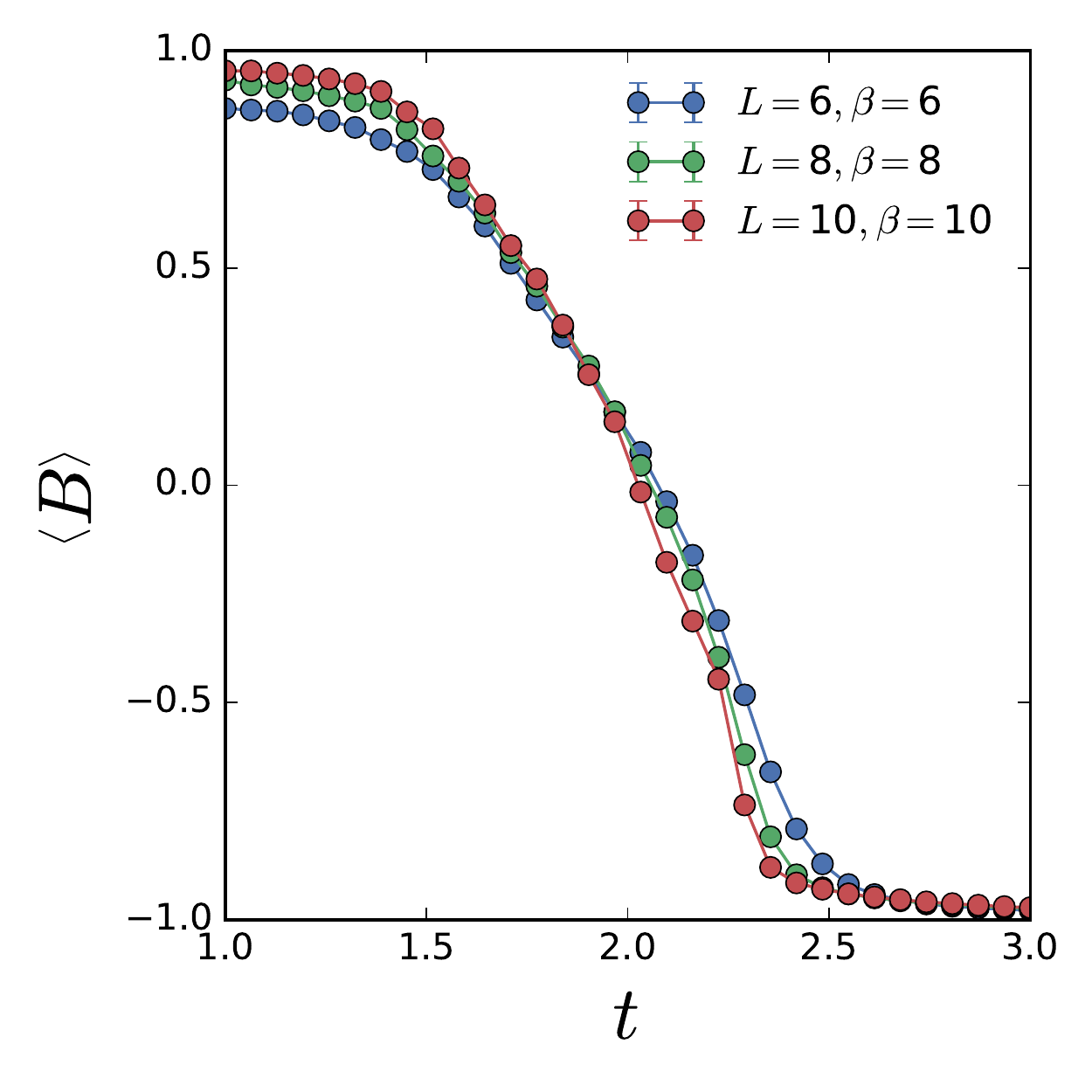}
		\caption{}\label{subfig:B_thop_bcs}	
		\hfill	
	\end{subfigure}
	\begin{subfigure}[t]{0.23\textwidth}
		\includegraphics[scale=0.35]{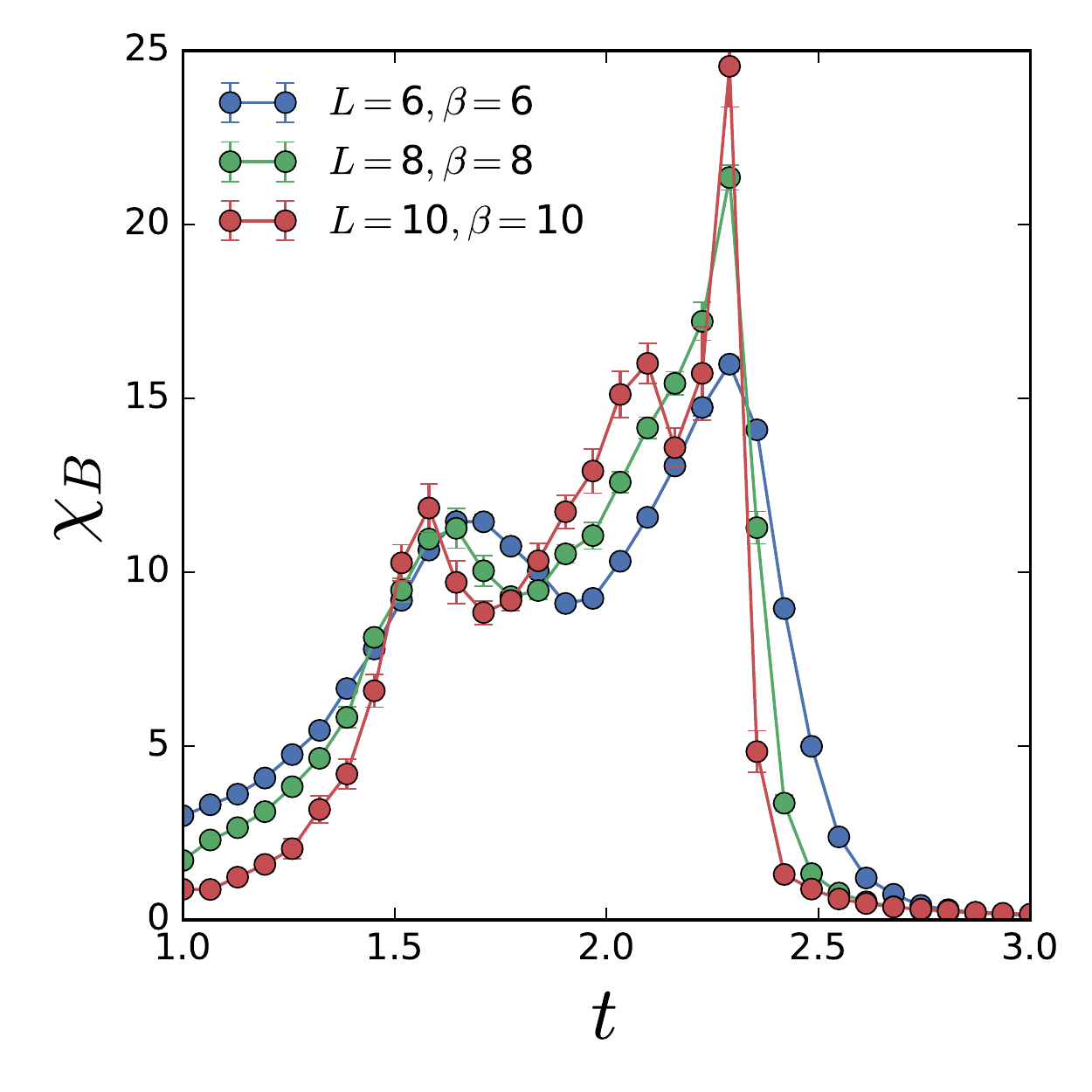}
		\caption{}\label{subfig:chiB_thop_bcs}		
	\end{subfigure}
	\hfill
	\begin{subfigure}[t]{0.23\textwidth}
		\includegraphics[scale=0.35]{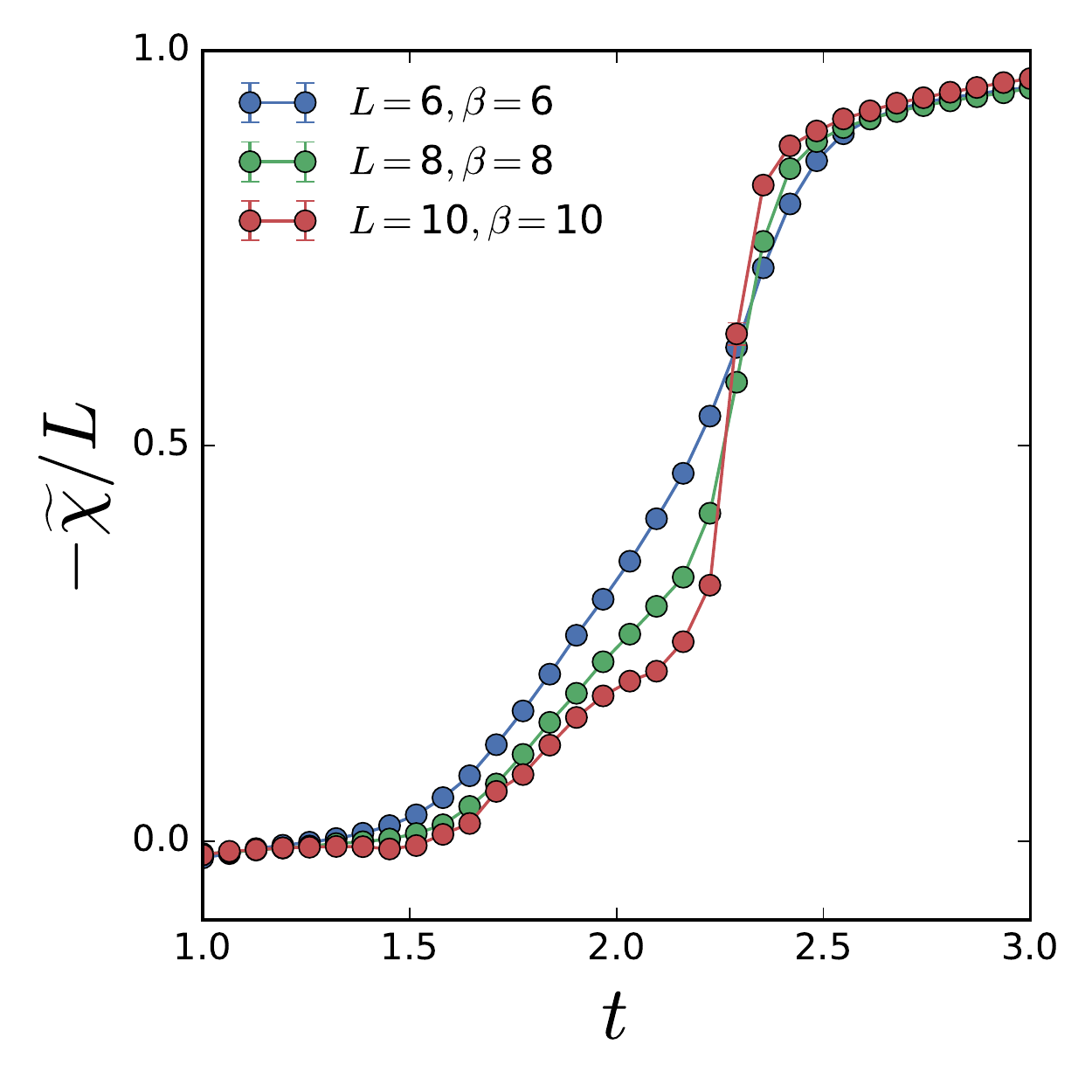}
		\caption{}\label{subfig:rhos_scling_thop_bcs}		
	\end{subfigure}
	\caption {The evolution from a deconfined BCS state to a deconfined Dirac phase with increasing $t$
		in the weak coupling regime $J=0.3$ $h=0.1$. 
		(a) The average Ising magnetic flux $\langle B \rangle$ which goes from $+1$ (or zero flux) deep in the deconfined BCS phase at small $t$
		to $-1$ (or $\pi$-flux) deep in the deconfined Dirac limit at large $t$. 
		(b) The Ising magnetic flux susceptibility $\chi_B$ exhibits two peaks as a function of $t$ indicating two distinct
		confinement transitions, with an intermediate-$t$ phase which is a confined BEC.
		(c) Scaled diamagnetic susceptibility $- \widetilde{\chi}/L$ (see text) as a function of $t$. 
		This goes to unity independent of $L$ in the Dirac phase at large $t$.
	}
\end{figure}

It is therefore convenient to analyze the QMC data in terms of the scaled 
susceptibility 
\begin{equation}
\widetilde\chi \equiv 4\pi\chi(q=2\pi/L)/t,
\end{equation}
and we plot in Fig.~\ref{subfig:rhos_scling_thop}
the $t$ dependence of $- \widetilde{\chi}/L$. 
We see clear evidence that $- \widetilde{\chi}/L$ approaches unity in the large $t$ limit.

We can also use the diamagnetic susceptibility as a diagnostic for superconductivity.
In a superconducting phase, the superfluid stiffness $\rho_s \neq 0$, and thus 
the diamagnetic susceptibility $\chi(q = 2\pi/L) \approx - L^2 \rho_s/4\pi^2$.
(We note in passing that, from this perspective, we can think our preceding analysis of the Dirac phase
in terms of $\rho_s \sim 1/L$; see ref.~\cite{LudwigFisherDirac}).

Before presenting the numerical results, we note that, as discussed before, for $\mu=0$ our model has an enlarged $SU(2)$ pseudo-spin symmetry, which transforms between superconductivity and charge density wave (CDW) order. The superfluid stiffness is then related to the spin stiffness
 of the SU(2) symmetry. Therefore, unlike the U(1) case for $\mu > 0$ where there is finite superfluid stiffness below the  BKT $T_c$, in the SU(2) case for $\mu = 0$, the
 spin stiffness is non-zero only at strictly zero temperature. Nevertheless, we can still probe zero temperature order by properly scaling both the linear system size $L$ and the inverse temperature $\beta$. 

To see superconductivity in the confined BEC phase, we plot in Fig.~\ref{subfig:rhos_thop}
$- \pi\widetilde{\chi}/L^2$ as a function of $t$, and find that in the small $t$ regime
the results are indeed independent of $L$ and $\beta$.

We further study the superconducting transition in Fig.~\ref{subfig:pairing_thop}, where we depict the static zero momentum  s-wave pairing susceptibility $\tilde{P}_{SC}=P_{SC}(q=0,i\omega_m=0)$, which serves as an order parameter for superconductivity. We note that, according to Mermin-Wagner theorem continuous symmetries can not be spontaneously broken in two dimensions at finite temperature. Therefore, we must  investigate the zero temperature ordering by scaling both the linear system size $L$ and inverse temperature $\beta$.
	
{\bf Transition between deconfined Dirac and confined Superfluid:}
Indeed, for small $t$, in the superconducting phase, the pairing susceptibility is finite. As $t$ increases the pairing susceptibility vanishes {\em continuously} at a critical coupling $t_c\approx 1.8(1)$ and remains zero throughout the non-superconducting Dirac phase.  Interestingly, the critical coupling $t_c$, reveled by the vanishing of the superconducting order parameter, appears to coincides with the peak in the magnetic field susceptibility $\chi_B$ which marks the confinement transition. 

If indeed, both the fermions and the Ising gauge field are critical,  the putative phase transition is expected to belong to a novel universality class which is distinct from either the usual chiral symmetry breaking Gross-Neveu \cite{HerbutGross} universality class or the confinement transition of the Ising lattice gauge theory (3D classical Ising model universality class). Determining the ultimate fate of this transition would require a refined  finite size scaling analysis that we will present in \cite{GazitInPre}.

{\bf Weak Coupling:} 
Finally, we look at the phase diagram at $\mu = 0$ at fixed $h/J < 1$, the weak coupling regime for the gauge theory.
We choose $J=0.3, h=0.1$ and examine how the system evolves as a function of the fermion hopping $t$.
In Fig.~\ref{subfig:B_thop_bcs} we plot the average Ising magnetic flux $\langle B \rangle$ which is a monotonically decreasing
function of $t$ with following limiting behavior. At small $t$, we find that  $\langle B \rangle \to +1$ or zero flux, characteristic
of a system deep in the deconfined phase, where the fermions form a BCS superconducting state. At the opposite
extreme of large $t$, $\langle B \rangle \to -1$ characteristic of a deconfined phase with a spontaneously
generated $\pi$-flux per plaquette and emergent Dirac nodes (see below). 

The system evolves between the deconfined BCS and deconfined Dirac phases in an interesting way.
Instead of showing a direct transition between the two, there is an intermediate phase. We see this 
in Fig.~\ref{subfig:chiB_thop_bcs} where the Ising magnetic flux susceptibility $\chi_B$ shows two peaks as a function of $t$ 
indicating two distinct confinement transitions. This implies a confined phase intermediate between 
the two deconfined phases, and by the arguments in the preceding Section, we identify this as
a confined BEC.

Finally, we show the existence of emergent Dirac excitations in the $\pi$-flux phase at large $t$, by analyzing the 
diamagnetic susceptibility in Fig.~\ref{subfig:rhos_scling_thop_bcs}. We see that at large $t$
$- \widetilde{\chi}/L \to 1$ independent of $L$, which is direct evidence for Dirac nodes.
The confined BEC to deconfined Dirac phase transition at weak coupling should be
in an interesting new universality class, as already mentioned in the case of a similar transition
in the strong coupling regime.

	\section{A particle-hole transformation between the even and odd  ILGT sectors}
	
	Before we conclude, in this section we derive an exact mapping between the "even"
	(without background charge) and the "odd" (with one background charge per unit cell) sectors of the 
	ILGT \cite{MoessnerRVB,SenthilZ2} at half filling. The mapping is generated by applying a partial particle hole transformation only on the spin down (without loss of generality) fermion \cite{NagaokaPseudoSpin}. This is the same transformation that in the context of a Hubbard model on a bipartite lattice exchanges repulsive and attractive interactions\cite{AuerbachBook}. At half filling, the Hamiltonian is invariant whereas the Ising Gauss` law obtains a non trivial minus sign factor which transforms the theory to its odd sector,
	\begin{equation}
		\prod_{b\in +_\sr} \sigma^x_{b}(-1)^{n^f_\sr}=-1
		\label{eq:constraint_odd}
	\end{equation}
Importantly, the partial particle hole transformation also maps the fermion number operator to the $z$ component of the spin operator $S^z=n_\uparrow-n_\downarrow$.

\begin{figure}[t!]
		\includegraphics[scale=0.4]{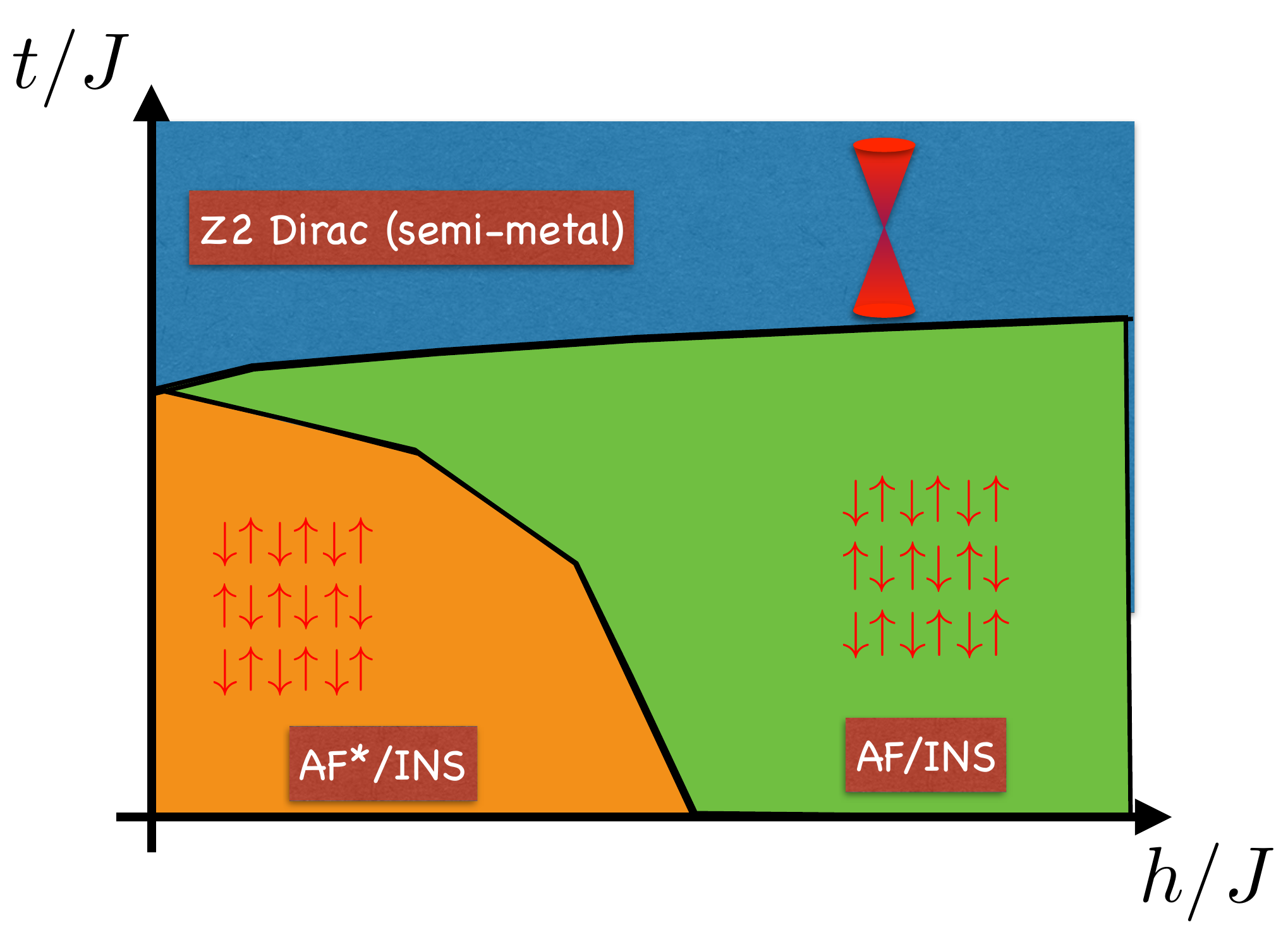}
		\caption{Schematic $T=0$ phase diagram of the "odd`` sector at half filling with $\mu=0$.}
		\label{fig:dual_phase_diag}
	\end{figure}

Interestingly, this mapping allows us to determine the odd sector phase diagram based on our analysis of the "even" sector in the previous section, see Fig.~\ref{fig:dual_phase_diag}. Due to the $SU(2)$ pseudospin symmetry, superconductivity in the even sector is degenerate with  CDW order and hence the partial particle hole symmetry maps  CDW order to an ordered antiferromagnetic (AF) spin density wave (SDW). As in the even sector, the deconfined state sustains non trivial fractional excitations and we denote it by AF$^*$ following the notation of \cite{SenthilZ2}. As a consequence, the AF states in the confined and deconfined phases are distinct and the transition between them is  an ILGT confinement transition. Within this model, the magnetically ordered phases are insulating while the deconfined Dirac theory is a semimetal, with gapless spin and charge excitations.

\section{Conclusions}

Our main conclusions are already summarized at the end of introduction, so we end with 
some remarks on the methodological progress and open questions.

First let us comment on the QMC sign problem, known to plague most fermion models at arbitrary density, 
with the well known exception of the attractive Hubbard model \cite{HirschDQMC}.
In recent years, sign-problem free QMC algorithms have been devised for many other interesting fermion problems.
They include models relevant for high energy physics \cite{ChandrasekharanSignFree}, Majorana fermion
based algorithms \cite{HongMajorana}, fermions coupled to nematic fluctuations \cite{SchattnerNematic} antiferromagnetic spin fluctuations \cite{ErezAntiferro,SchattnerAntiferro} and multi-band models relevant for pnictides \cite{LiDQMC,dumitrescuFeSe}

The sign-problem free algorithm introduced in this paper is, to the best of our knowledge, the first one for 
charged fermions couple to gauge theories. A different problem that we faced above, and were able to overcome, is the
``zero problem". Here the neglect of a class of configurations which have vanishing weight (due to a symmetry)
leads to a bias in the evaluation of certain observables. Our solution of the ``zero problem" problem by 
using an extended configuration space may be of some general interest for QMC simulations.

Note that we work in the ``even sector'' of the gauge theory with no background $\mathbb{Z}_2$ charge.
We do not know of general sign-problem free QMC algorithm for the odd sector \cite{SenthilZ2},
except for the special case of half-filing \cite{GazitInPre}. This sector is 
relevant for Mott insulators, with a background of one fermion per site, and for
describing d-wave superconductivity upon doping.

Our focus in this paper has been on understanding the phases and the phase diagram
of fermions coupled to $\mathbb{Z}_2$ gauge theory in $(2+1)$D, a basic problem on which there has been 
essentially no prior work, that we are aware of. The next step is to gain insight into
the universal critical behavior at the quantum phase transitions. This
requires a finite size scaling analysis of QMC data, an important task that we leave for a future paper. 

At generic filling, the fermions undergo a BCS to BEC crossover in the fermion sector that rides on top of an underlying 
deconfined to confined phase transition for the gauge fields. We expect this transition to be in the same (3D Ising) 
universality class as confinement in pure $\mathbb{Z}_2$ gauge theory, since the fermions are gapped out. 
In the half-filled case, we find an additional phase with emergent Dirac excitations for large fermion hopping.  
Our results on this novel deconfined Dirac phase raise many questions that are worthy of further study. 
These include: understanding the universality class of the confined BEC to deconfined Dirac phase transition;
understanding the intermediate confined BEC phase between the deconfined BCS and Dirac phases at weak coupling,
and how it pinches off to give a Fermi surface area changing transition between two deconfined phases without
any symmetry breaking.

{\em Note added} : After finalizing the results of this paper we became aware of a QMC study of a related interesting  model\cite{TraunUnPub}, in which the Ising Gauss' law is not explicitly enforced. Understanding the precise relation between these works is left to the future.
	
\section{Acknowldgements}
We would like to thank Subir Sachdev and T. Senthil for discussions. 
SG received support from the Simons Investigators Program, the California Institute of Quantum Emulation and the Templeton Foundation, AV acknowledges support from the Templeton Foundation and a Simons Investigator Award.
MR would like to acknowledge support from NSF DMR-1410364 and the hospitality of the Condensed Matter Theory Group at Berkeley
in Fall 2015. This research was done using resources provided by the Open Science Grid \cite{OSG1,OSG2}, which is supported by the National Science Foundation and used the Extreme Science and Engineering Discovery Environment \cite{xsede} (XSEDE), which is supported by National Science Foundation grant number ACI-1053575. 

\bibliography{Z2LGT}{}

\appendix

\section{DQMC - Path integral formulation}
\label{app:DQMC}
In this section we map the two dimensional quantum problem to a three dimensional classical statistical mechanics model. This is done by rewriting  the grand canonical partition function $\mathcal{Z}(\beta,\mu)$ ( Eq.~\eqref{eq:partition_function}) in terms of an imaginary time path integral.
The procedure closely follows the standard QMC methods with the exception  that in our case we must also incorporate the constraint (Eq.~\eqref{eq:constraint}). 

We define a projection operator, $\hat{P}=\prod_\sr \hat{P}_\sr$, which imposes the constraint in Eq.~\ref{eq:constraint} at each site $\sr$ \cite{SenthilZ2},
\begin{equation}
	\hat{P_\sr}=\frac{1}{2}\left(1+\prod_{\eta=\pm \hat{e}_x,\hat{e}_y} \sigma^x_{\sr+\eta}(-1)^{n_\sr}\right)
	\label{eq:proj_op}
\end{equation}
In the path integral formulation, we will use an equivalent expression using a discrete Lagrange multiplier,
\begin{equation}
	\hat{P}_r=\sum_{\lambda_\sr=\pm 1} \hat{P}_{\lambda_\sr} =\frac{1}{2}\sum_{\lambda_\sr=\pm 1} e^{i\frac{\pi}{2}(1-\lambda_\sr) \left(\sum_{\eta} \frac{\left(1-\sigma^x_{\sr+\eta}\right)}{2} +n_\sr\right)}
\end{equation}
The Ising gauge fields $\lambda_\sr$ are identified with the temporal gauge field in the Lagrangian formulation of the ILGT \cite{KogutDuality}.

Next, we use a Trotter decomposition to write the thermal density matrix as $e^{-\beta \mathcal{H}}=\prod_{\tau=0}^{M-1}e^{-\epsilon \mathcal{H}}$ with $\epsilon=\beta/M$ and introduce resolution of the identities in the $\sigma^z$ basis, $\mathds{1}=\sum_{\sigma^z} \ket{\sigma^z}\bra{\sigma^z}$, between each imaginary time step,
\begin{equation}
	\begin{aligned}
		\mathcal{Z}(\beta,\mu)&=\sum_{\sigma^z_{r,\eta}} Tr_f \left[ \bra{\sigma^z_{\tau=0}}  \hat{P} e^{-\epsilon\mathcal{H}}\ket{\sigma^z_{\tau=M-1}} \right. \\ \times& \bra{\sigma^z_{\tau=M-1}}e^{-\epsilon\mathcal{H}}\ket{\sigma^z_{\tau=M-2}}\dots\bra{\sigma^z_{\tau=1}} \left.e^{-\epsilon\mathcal{H}}\ket{\sigma^z_{\tau=0}}\right].
	\end{aligned}
	\label{eq:trotter}
\end{equation}
In the above equation we use a unified space time notation, such that $r=\{\sr_x,\sr_y,\tau\}$ and the temporal Ising gauge field is then, 
\begin{equation}
	\sigma^z_{r=\{\sr,\tau\},\unittau}=\begin{cases} 
		\lambda_\sr & \tau=M-1  \\
		1 & \text{else}
	\end{cases}
\end{equation}

We note that at finite temperature, the periodic boundary conditions along the imaginary time axis leads to a non trivial cycle. The temporal gauge field, therefore,  can not be completely eliminated.

Following standard techniques \cite{AssaadDQMC}, we can compute the matrix elements appearing in Eq.~\ref{eq:trotter} to order $\mathcal{O}(\epsilon^2)$.  We focus on the first term containing the projection operator ,
\begin{equation}
	\begin{aligned}
		\bra{\sigma^z_{\tau=0}} \hat{P}_{\lambda_\sr}e^{-\epsilon\mathcal{H}}\ket{\sigma^z_{\tau=M-1}} =  \\ e^{i \frac{\pi}{2}\sum_{\sr}(1-\lambda_\sr) n_\sr} e^{\epsilon\mathcal{H}_f\left[\sigma^z_{\tau=M-1}\right]} \times W_{\sigma^z_{\tau=M-1 }}+\mathcal{O}(\epsilon^2)
	\end{aligned}
\end{equation}
The imaginary time depended fermion Hamiltonian is given by,
\begin{equation}
	\begin{aligned}
		\mathcal{H}_f\left[\sigma^z_{\tau=M-1}\right]&=\sum_{\sr,\eta,\alpha} \mathcal{K}_{\sr,\sr+\eta}c^\dagger_{\sr,\alpha}c_{\sr+\eta,\alpha} + h.c. \\ 
		& - \mu\sum_{\sr,\alpha} c^\dagger_{\sr,\alpha}c_{\sr,\alpha}
	\end{aligned}
\end{equation}
Explicitly, the kernel matrix equals $\mathcal{K}_{\sr,\sr+\eta}({\tau=M-1})=-t\mathcal{\sigma}^z_{\sr,\sr+\eta}$.

The Boltzmann weight associated with each gauge field configuration, $W_{\sigma^z_{\tau=M-1 }}=e^{S_{\sigma^z_{\tau=M-1 }}}$, is  given by the classical action, 
\begin{equation}
	\begin{aligned}
		S_{\sigma^z_{\tau=M-1 }}&= \gamma \sum_{\sr,\eta=\unitx,\unity} \prod_{b\in \square_{\sr,\tau=M-1,\eta} }\sigma^z_b \\
		&+ \epsilon J \sum_{\sr} \prod_{b\in \square_{\sr,\tau=M-1,\hat{e}_\tau}} \sigma^z_b .
	\end{aligned}			
	\label{eq:IGFaction}
\end{equation}
where  $\gamma=-\frac{1}{2}\log{\left(\tanh \epsilon h\right)}$. In the first term, the plaquette $\square_{\sr,\tau,\eta} $ is a spatio-temporal plaquette defined by the space time point $r=\{\sr,\tau\}$ and the direction $\eta$. For instance, $\square_{\sr,\tau,\unitx}$ corresponds to the set of bonds  $b=\{\{r,\unitx\},\{r,\unittau\},\{r+\unity,\unittau\},\{r+\unittau,\unitx\}\}$. In the second term, the plaquette is a planar plaquette defined similarly to Ising magnetic flux term of the Hamiltonian.

For the rest of the time slices the Boltzmann weight is readily evaluated in a similar manner. The temporal gauge field in this case is trivial $\sigma^z_{\sr, \tau \ne M-1,\unittau}=1$.

The fermionic weight amounts to  tracing over a product of quadratic fermion propagators\cite{AssaadDQMC},
\begin{equation}
	\begin{aligned}
		w_f(\{\sigma^z_{r,\eta}\}) &=Tr_f\left[\ e^{i \frac{\pi}{2}\sum_{\sr}(1-\lambda_\sr) n_\sr^\uparrow} \prod_\tau e^{-\epsilon\mathcal{H}[\sigma^z_\tau]}\right] \\
		&=\det{\left(I+P[\lambda_\sr]\prod_\tau e^{-\epsilon \mathcal{K}({\tau})}\right)}
	\end{aligned}
	\label{eq:f_weight}
\end{equation}  
Here, the projector is manifested by the diagonal matrix  $P[\lambda_\sr]$  with elements  $P_{\sr,\sr}=\lambda_\sr$.
For future convince we also  define the equal time single particle Green's function, which for a given gauge field configuration equals,
\begin{equation}
	G=\left(I+P[\lambda_\sr]\prod_\tau e^{-\epsilon \mathcal{K}({\tau})}\right)^{-1}
	\label{eq:green_conf}
\end{equation}
The  total weight of the fermionic sector is then a product over the spin up and spin down sector,
\begin{equation}
	W_f=w_f^2 =\left[\det{\left(I+P[\lambda_\sr]\prod_\tau e^{-\epsilon \mathcal{K}({\tau})}\right) }\right]^2
\end{equation}

Since both determinants are real, the weight in strictly non negative and hence free from the numerical sign problem.

\section{Particle-Hole symmetry and zero modes}
\label{app:PH}
At zero chemical potential, $\mu=0$, both the Hamiltonian {\em and } the constraint are symmetric under the particle hole (PH) transformation $\mathcal{C}$, defined by,
\begin{equation}
	c_{\sr,\alpha}\rightarrow\begin{cases}
		c_{\sr,\alpha}^\dagger & \sr\in A \\
		-c_{\sr,\alpha}^\dagger & \sr\in B
	\end{cases}
	\label{eq:PH}
\end{equation}

where the $A$ and $B$ sub-lattices correspond to the usual checkered board division of the square lattice (or more generally any bipartite lattice) to two disconnected sub-lattices. PH symmetry has a dramatic effect on the fermionic configuration weight Eq.~\eqref{eq:f_weight}. To see that, we apply the PH transportation, without loss of generality, only on the spin up, $\alpha=\uparrow$, sector of the fermionic weight in Eq.~\eqref{eq:f_weight}. We denote this operator by $\mathcal{C}_{\alpha=\uparrow}$. Since the Hamiltonian is symmetric under PH,  the only non-trivial transformation is due to the constraint. Explicitly, 
\begin{equation}
	\begin{aligned}
		\mathcal{C}_\uparrow^\dagger e^{i \frac{\pi}{2}\sum_{\sr}(1-\lambda_\sr) n_\sr^\uparrow}\mathcal{C}_\uparrow=e^{i \frac{\pi}{2}\sum_{\sr}(1-\lambda_\sr)} e^{i \frac{\pi}{2}\sum_{\sr}(1-\lambda_\sr) n_\sr^\uparrow}
	\end{aligned}
\end{equation} 

As a direct consequence, if the parity of the temporal Ising gauge field $\mathcal{P}_\lambda=\prod_{\sr}\lambda_\sr$ is odd, $\mathcal{P}_\lambda=-1$, the fermion weight obeys, $W_f^\uparrow(\{\sigma_{r,\eta}\}|\mathcal{P}_\lambda=-1) =-W_f^\uparrow(\{\sigma_{r,\eta}\}|\mathcal{P}_\lambda=-1)$ and hence it must vanish. 

The vanishing of the fermion determinant indicates on the presence of a {\em finite} temperature fermionic zero mode. This result is surprising, since due to the anti-periodic boundary conditions along the imaginary time axis the lowest Matsubara frequency of fermions is non vanishing, $T\pi$, and hence can not sustain poles on the real frequency axis.  
	
We note that in our case the projection operator couples the temporal Ising gauge field to the density operator and acts as an effective  complex chemical potential. In the odd sector, this effect shifts the lowest Matsubara frequency by $T\pi$ down to zero and gives rise to a zero mode.  

Naively, the above result does not affect the Monte Carlo sampling since it merely leads to a vanishing probability for configurations with odd parity. However, it gives rise to  a systematic bias in computing expectation values of observables that are not symmetric under PH transformation of a single spin flavor,   $\mathcal{C}_{\uparrow/\downarrow}$. 

To address this problem, we introduce an extended configuration space which enables us to sample the contribution of the odd sector. For concreteness, we consider the pairing susceptibility.  The derivation can be readily generalized to other observables. 

The expectation value of the equal-time paring susceptibility is given by,
\begin{equation}
	\begin{aligned}
		P_{SC}(\sr,\sr,\tau=0)&=\frac{1}{\mathcal{Z}}\sum_{\sigma^z_{\lambda,\eta}}W_{\sigma^z_{\lambda,\eta}}Tr_f\left[\ e^{i \frac{\pi}{2}\sum_{\sr}(1-\lambda_\sr) n_\sr} \right.\\
		&\times\left. \prod_\tau e^{-\epsilon\mathcal{H}[\sigma^z_\tau]}b_\sr^\dagger b_\sr\right]
	\end{aligned}
\end{equation} 
where $b_\sr^{\dagger}=c^\dagger_{\sr,\uparrow}c^\dagger_{\sr,\downarrow}$. The above can be readily evaluated using Wick's theorem.
\begin{equation}
	P_S(\sr,\sr,\tau=0)=\frac{1}{\mathcal{Z}}\sum_{\sigma^z_{\lambda,\eta}}W_{\sigma^z_{\lambda,\eta}} W_{f,\sigma^z_{\lambda,\eta}} G_{\sr,\sr}^2(\sigma^z_{\lambda,\eta}) 
\end{equation}

As shown before, In the odd sector, the fermionic weight vanishes and hence the Green's function $G$ diverges. The product, however, is {\em finite}.  One possible solution for circumventing the ratio of zeros problem is to perform the MC simulation on a set of decreasing but finite chemical potentials.  The zero chemical potential result can  then be obtained by extrapolation. This method significantly complicates the computations and the fitting procedure introduces additional numerical errors. 

In the following we will introduce a simple solution that does not require breaking of PH symmetry. We first artificially break PH symmetry by introducing a small but finite chemical potential $\mu>0$,  rendering our calculation regular. In the last step we will recover the zero chemical potential result by taking the limit  $\mu\to0$ {\em analytically}.  
The odd sector contribution involve the {\em finite} product,
\begin{equation}
	G^H=\det{(G^{-1})} G
	\label{eq:adjugate}
\end{equation}
In the above equation, we identified the product with the adjugate matrix \cite{adjugate}.  

To evaluate the product we must eliminate the singularity. This can be achieved by a singular value decomposition (SVD) analysis \cite{adjugate}. Explicitly, we write $G^{-1}=UDV^T$ where $U,V$ are orthogonal matrices and $D$ is diagonal matrix with positive entries  $D_{i,i}=d_i$ known as the singular values. We substitute the SVD decomposition in Eq.~\eqref{eq:adjugate} and obtain,
\begin{equation}
	G^H=\det(U)\det(V)\det(D)VD^{-1}U^T
\end{equation}
In the odd sector, one of the singular values , ${d}_k$, vanishes in the limit of $\mu\to0$.  First we isolate the vanishing singular value $\det(D)=(\prod_{i\ne k} d_i ) \times d_k$. Now we can cancel the singularity appearing in $D^{-1}$,
\begin{equation}
	\lim_{\mu\to0}(d_k D^{-1})_{\ell,\ell}= \lim_{\mu\to0} \frac{d_k}{d_\ell} = \begin{cases}
		1 & \ell =k \\
		0 & \text{else}
	\end{cases}
\end{equation}
Finally we obtain,
\begin{equation}
	G^H=\det(U)\det(V)(\prod_{i\ne k} d_i ) v_{k}u_{k}^T 
	\label{eq:GH_nonsing}
\end{equation}
where  $v_k(u_k)$ correspond to the $k$'th column of the matrix $V(U)$.

The above analysis suggests the following Monte Carlo sampling scheme. We consider an extended configuration space $\tilde{\mathcal{Z}}=\mathcal{Z}_{\text{even}}+\mathcal{Z}_{\text{odd}}$. The configuration weight and Green's function of the even sector, $\mathcal{P}_\lambda=1$ are the same as the ones given in Eq.~\eqref{eq:f_weight} and Eq.~\eqref{eq:green_conf} respectively. For the odd sector, $\mathcal{P}_\lambda=-1$ we used Eq.\eqref{eq:GH_nonsing} te redefine both the configuration weight and the Green's function for the odd sector. Explicitly, 
\begin{equation}
	w_f^{\text{odd}} =\prod_{i\ne k} d_i ,\quad
	G^{\text{odd}}= v_{k}u_{k}^T
\end{equation} 
Since we sample with respect to an extended configuration, $\tilde{\mathcal{Z}}$ , we must use reweighting to correctly compute expectation values. This is readily achieved by sampling the fraction of the even sector configurations $\av{\delta_{\text{even}}}_{\tilde{\mathcal{Z}}}=\frac{\mathcal{Z}_{\text{even}}}{\tilde{\mathcal{Z}}}$, such that,
\begin{equation}
	\av{\mathcal{O}}_{\mathcal{Z}}=\frac{\av{\mathcal{O}}_{\tilde{\mathcal{Z}}}}{\av{\delta_{\text{even}}}_{\tilde{\mathcal{Z}}}}
\end{equation}

We note that the above scheme does not modify significantly the usual DQMC algorithm since the SVD decomposition is available as part of the stabilization scheme of DQMC \cite{AssaadDQMC}. 

\section{Updating scheme}
\label{app:worm}
To evaluate the partition function in Eq.\eqref{eq:partition_function}, we must devise an efficient scheme for sampling the configuration space $\{\sigma^z_{r,\eta}\}$. To achieve that, we use both a local updating approach \cite{AssaadDQMC} and a global updating strategy inspired  by the worm algorithm (WA) \cite{wormAlg,fabinclusterU1}.

The local updates, involve single spin flip of  the Ising gauge fields. Both the temporal and spatial updates can be performed efficiently using a low rank (rank one in the case of the temporal link and rank two in the case of the spatial link) updating of the determinant in Eq.~\eqref{eq:f_weight} and the corresponding Green's function. 

Empirically, we found that using solely local updates does not lead to a sufficiently short MC correlation time. This  effect is prominent near the critical point where the dynamics is critically slowed down \cite{BinderMCBook} due to the diverging  correlation length. To tackle this problem, we introduce an additional MC move based on the highly efficient WA.
\begin{figure}
	\includegraphics[scale=0.5]{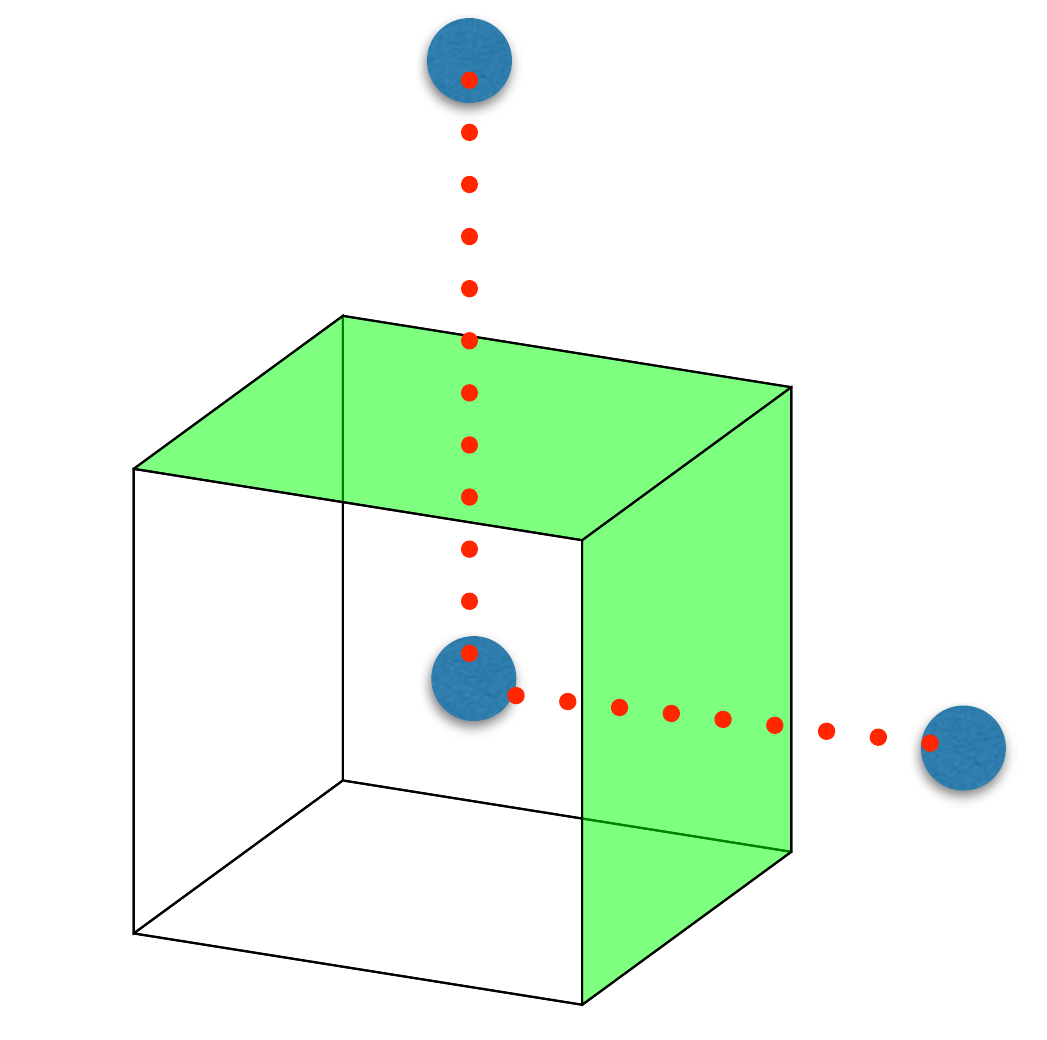}
	\caption{Dual loop representation of the classical Ising lattice gauge theory. The loops are constructed by connecting dual lattice point (blue circles) that share a frustrated plaquette (green facets).  }
	\label{fig:dual_loop}
\end{figure}

We reformulate the Ising gauge field sector of the action in a dual closed loop representation \cite{SavitDuality}. We note that this mapping is used in deriving the classical statistical mechanics duality between the classical Ising gauge theory and Ising model in three dimensions. 

The closed loop configurations are constructed as follows. We first identify all frustrated space-time plaquettes  $\square_r$  satisfying $\prod_{b\in \square_{r}} \sigma^z_b =-1$. We then draw a line connecting the two neighboring sites of the dual three dimensional cubic lattice that share the frustrated plaquettes, see Fig.~\ref{fig:dual_loop}. Since the ILGT is free of magnetic monopoles, the net flux through each elementary cube must be even. Therefore, the number of dual lattice lines emanating each dual lattice the site (located at the center of the direct lattice cube) must be also even. This constraint enforces the lines to form a closed loop configuration \cite{wormAlg}. Periodic boundary conditions along the spatial and temporal directions give rise to an additional constraint. The net flux along each plane must be even. This is in contrast to the closed loop representation of the classical  Ising model, for which the total parity of the loops can fluctuate.   A similar construction was proposed in the context of $U(1)$ gauge theory  \cite{fabinclusterU1} .

The closed loop ensemble can be efficiently sampled using the WA, where the worm head flips the flux through the plaquette and generates arbitrary flux tubes. The loop fugacity is anisotropic and is determined according the Ising gauge field action 
(Eq.~\eqref{eq:IGFaction} ). During the loop update we ignore the fermionic weight $W_f$. After the loop is closed it is reintroduced in the acceptance probability of the entire move. In case that the move was accepted we translate the flux configuration to a gauge configuration. In the temporal gauge, this  can be done uniquely up to a global gauge freedom in each space-time direction which is drawn randomly \cite{LGTbook}.

\section{Benchmarking against exact diagonalization}
\begin{figure}[t!]
	\includegraphics[scale=0.5]{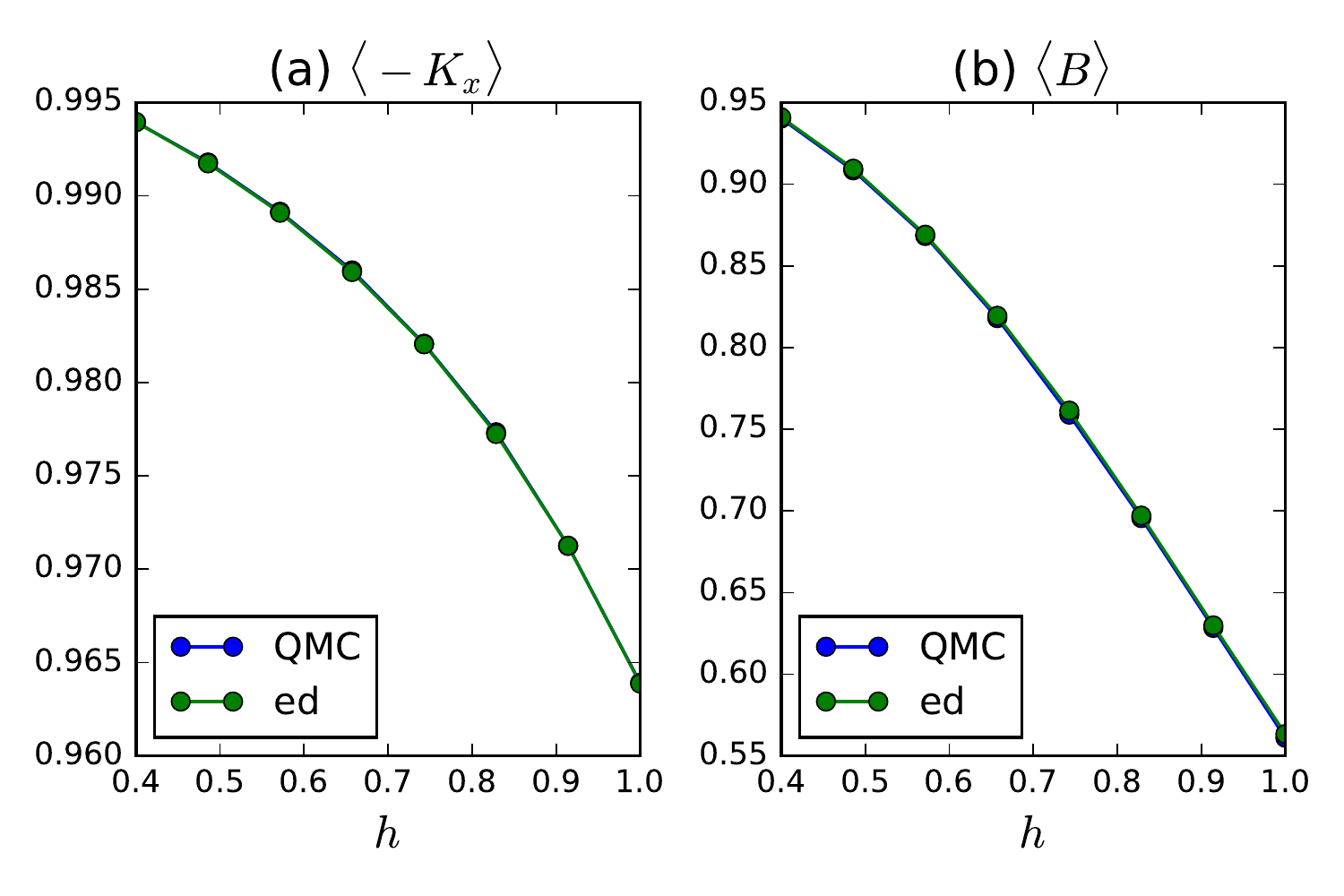}
	\caption{Comparison between exact diagonalization (ed) and QMC results for  (a) kinetic energy $\langle -K_x \rangle$ and (b) Ising magnetic flux $\langle B \rangle$. 	 }
	\label{fig:ed}
\end{figure}

\begin{figure}[t!]
	\includegraphics[scale=0.5]{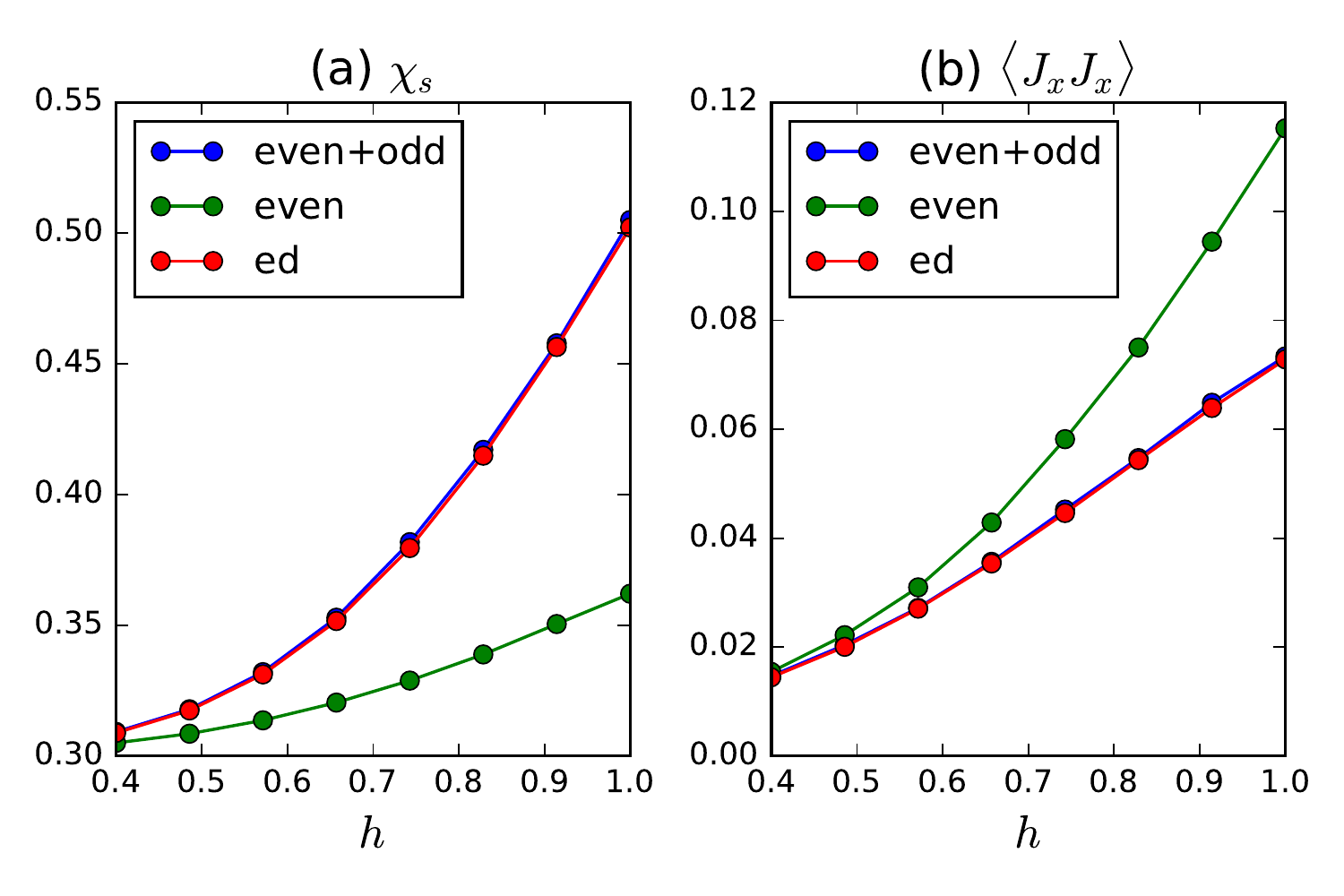}
	\caption{Comparison between exact diagonalization (ed) and the even and odd contributions to the (a) pairing susceptibility $P_{SC}$ and (b) current current correlation function $\langle J_x J_x \rangle$. The even sector contribution (green curve) does not match the ED result (red curve). The difference is exactly compensated by the odd sector contribution (blue curve).  }
	\label{fig:ed_mu}
\end{figure}
We verify the correctness of our numerical scheme by comparing the QMC results with exact diagonalization (ED) on a small system with $L=2$. As concrete microscopic parameters we take $\beta=2,t=1,J=1,\mu=0$ and consider a set of eight evenly spaced points in  $h\in[0.4,1.0]$. The ED is performed by diagonalizing the Hamiltonian  Eq.~\eqref{eq:hamiltonian} restricted to the subspace of physical states satisfying the constraint in Eq.~\eqref{eq:constraint}.  

In Fig.~\ref{fig:ed} we compare the average kinetic energy, $\langle -K_x \rangle$ and the  average Ising magnetic flux $\langle B \rangle$. We find excellent agreement within the statistical error. In Fig.~\ref{fig:ed_mu} we consider the pairing susceptibility $P_{SC}(q=0,i\omega_m=0)$ and the current-current correlation function $\langle J_x(q=0,i\omega_m=0) J_x (q=0,i\omega_m=0)\rangle$. We note that both observable are not symmetric under $\mathcal{C}_\uparrow$. In the Figure we demonstrate that the procedure out lined in \ref{app:PH} precisely  compensate on the missing weight of the odd sector.

\section{The $\pi$-flux lattice}

\label{app:piflux}
The hopping Hamiltonian of the $\pi$-flux lattice is given by,
\begin{equation}
	\mathcal{H}=-\sum_{r,r'} t_{r,r'}c_r^\dagger c_{r'}+h.c.,
	\label{eq:piflux_H}
\end{equation} 
where $t_{r,r'}=t \left[(-1)^{r_x}\delta_{r,r+\unity}+\delta_{r,r+\unitx}\right]$.
To diagonalize the Hamiltonian we first double the unit cell, such that the total flux in the enlarged unit cell equals $2\pi$. We arrange the fermion operators belonging to each unit cell in a two component spinor,
\begin{equation}
	\Psi^\dagger_{R}=\{c^\dagger_{R,A},c^\dagger_{R,B}\}
\end{equation}  
where the sub lattice $\{R,A\}\,(\{R,B\})$ is defined by the set of lattice points $\{r_x,r_y\}=\{2n,m\}(\{2n+1,m\})$.
We now transform the Hamiltonian to momentum space, 
\begin{equation}
\mathcal{H}=-t\sum_{k} \Psi^\dagger(k) \begin{pmatrix}
-2\cos(k_y) & 1+e^{i 2 k_x} \\
1+e^{-i 2 k_x}  & 2\cos(k_y)
\end{pmatrix} \Psi(k) 
\end{equation}
where,
\begin{equation}
	c_{A/B}(k)=\sum_{R} e^{i R \cdot k} c_{R,{A/B}},
\end{equation} 
\begin{figure}[t]
	\includegraphics[scale=0.5]{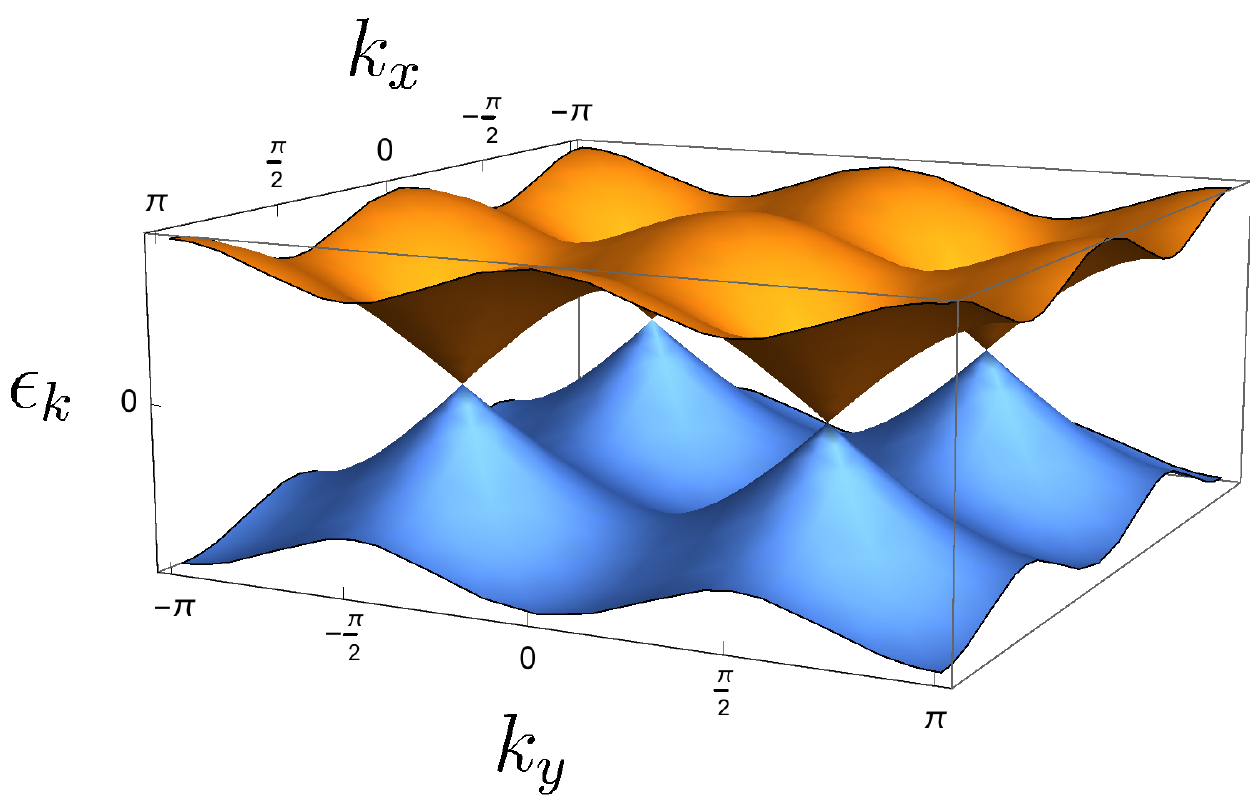}
	\caption{Dispersion relation of the $\pi$-flux lattice. The first Brillouin zone contains two Dirac nodes at $\vec{k}=\{\pi/2,\pm\pi/2\}$ }
	\label{fig:piflux_disp}
\end{figure}
and the first Brillouin zone is defined by the region $\pi/2\le k_x \le \pi/2$ and $\pi\le k_y \le \pi$.
It is convenient to express the $2\times2$ matrix kernel using the Pauli matrices $\vec{\sigma}$ as,
\begin{equation}
	\mathcal{H}=-t\sum_{k} \Psi^\dagger(k) \vec{d}_k\cdot \vec{\sigma}\Psi(k) 
\end{equation}
where $\vec{d}_k=\{1+\cos(2 k_x),\sin(2 k_x),-2\cos(k_y)\}$. 
The dispersion relation is then,
\begin{equation}
	\epsilon_k=\pm 2t\sqrt{\cos^2(k_y)+\cos^2(k_x)}
\end{equation}

which contains two Dirac nodes at $\vec{k}=\{\pi/2,\pm\pi/2\}$ with a Fermi velocity $v_F=2t$, see Fig.~\ref{fig:piflux_disp}.

\end{document}